\documentclass{aa}
\usepackage[varg]{txfonts}
\usepackage{graphicx}
\usepackage{txfonts}
\usepackage{natbib}
\usepackage{verbatim}
\usepackage[pdftex]{color}
\usepackage[dvipsnames]{xcolor}
\usepackage{hyperref}
\usepackage{ulem}
\usepackage{subfigure}
\hypersetup{
    colorlinks = true,
    citecolor ={blue}
}
\usepackage{mathrsfs}
\bibpunct{(}{)}{;}{a}{}{,} % to follow the A&A style

\usepackage{booktabs}
\definecolor{ultramarine}{rgb}{0.07, 0.1, 0.6} 
\definecolor{myblue}{rgb}{0.07, 0.2, 0.6} 
\definecolor{dopal}{rgb}{.70, .25, .05}

\begin{document}

\title{Structure and evolution of ultra-massive white dwarfs in general relativity
  \thanks{The cooling sequences are publicly available  at
   \url{http://evolgroup.fcaglp.unlp.edu.ar/TRACKS/tracks.html}}}

    %\href{ http://evolgroup.fcaglp.unlp.edu.ar/TRACKS/tracks.html}}}

%resulting from
%  single stellar evolution}

\author{Leandro G. Althaus\inst{1,2},  Mar\'ia E. Camisassa\inst{3}, Santiago Torres\inst{4,5}, Tiara Battich\inst{6}, Alejandro H. C\'orsico\inst{1,2}, Alberto Rebassa-Mansergas\inst{4,5}, Roberto Raddi\inst{4,5} }
\institute{Grupo de Evoluci\'on Estelar y Pulsaciones. 
           Facultad de Ciencias Astron\'omicas y Geof\'{\i}sicas, 
           Universidad Nacional de La Plata, 
           Paseo del Bosque s/n, 1900 
           La Plata, 
           Argentina
           \and
           IALP-CCT - CONICET
            \and
Applied Mathematics Department, University of Colorado, Boulder, CO 80309-0526, USA            \and
           Departament de F\'\i sica, 
           Universitat Polit\`ecnica de Catalunya, 
           c/Esteve Terrades 5, 
           08860 Castelldefels, 
           Spain
           \and
           Institute for Space Studies of Catalonia, 
           c/Gran Capita 2--4, 
           Edif. Nexus 104, 
           08034 Barcelona, 
           Spain
           \and
            Max-Planck-Institut f\"{u}r Astrophysik, Karl-Schwarzschild-Str. 1, 85748, Garching, Germany
                      }
\date{Received ; accepted }
\abstract{Ultra-massive white dwarfs ($M_{\star} \gtrsim 1.05 M_{\sun}$)
  are of utmost  importance in view of  the role they play  in type Ia
  supernovae explosions, merger events, the existence of high magnetic
  field  white  dwarfs,  and  the  physical  processes  in  the  Super
  Asymptotic  Giant  Branch phase.}   {We  present  the first  set  of
  constant  rest-mass ultra-massive  oxygen/neon  white dwarf  cooling
  tracks with masses  $M_{\star} > 1.29 M_{\sun}$ which  fully take into
  account the  effects of general  relativity on their  structural and
  evolutionary properties.}
{We have  computed the  full evolution sequences  of 1.29,  1.31, 1.33,
  1.35, and  1.369 $M_{\sun}$ white  dwarfs with the La  Plata stellar
  evolution code, {\tt LPCODE}. For  this work, the standard equations
  of stellar structure and evolution have been modified to include the
  effects  of general  relativity.   Specifically,  the fully  general
  relativistic partial differential  equations governing the evolution
  of a  spherically symmetric star are  solved in a way  they resemble
  the  standard   Newtonian  equations   of  stellar   structure.  For
  comparison purposes, the  same sequences have been  computed but for
  the Newtonian case.}
{According to our calculations, the evolutionary properties of the most
  massive  white dwarfs  are strongly  modified by  general relativity
  effects.  In  particular, the  resulting stellar radius  is markedly
  smaller in the general relativistic case, being up to 25$\%$ smaller
  than  predicted by  the  Newtonian treatment  for  the more  massive
  ones.  We  find that  oxygen/neon  white  dwarfs  more massive  than  1.369
  $M_{\sun}$ become  gravitationally unstable with respect to general
  relativity  effects. When  core chemical  distribution due  to phase
  separation on crystallization is considered, such instability occurs
  at  somewhat  lower stellar  masses,  $\gtrsim  1.36 M_{\sun}$.   In
  addition, cooling times  for the most massive  white dwarf sequences
  result in about  a factor of two smaller than  in the Newtonian case
  at advanced stages of evolution.   Finally, a sample of white dwarfs
  has  been  identified as  ideal  candidates  to test  these  general
  relativistic effects.}
{%We have computed the evolution of ultra-massive white dwarfs by fully taking into account the general relativity effects. 
We conclude
that the general relativity effects should be taken into account for an accurate assessment of the
structural and evolutionary properties of the most massive white dwarfs. These new ultra-massive white dwarf models constitute a considerable improvement over those computed in the framework of the standard Newtonian theory of stellar interiors. }

\titlerunning{Relativistic ultra-massive white dwarfs}
\authorrunning{Althaus et al.}

\maketitle

\section{Introduction}
\label{introduction}

White  dwarf  stars   are  the  most  common  end   point  of  stellar
evolution.  Therefore, these  old  stellar  remnants contain  valuable
information on  the stellar evolution  theory, the kinematics  and the
star  formation  history of  our  Galaxy,  and  the ultimate  fate  of
planetary                    systems                   \citep[see][for
  reviews]{2008ARA&A..46..157W,2010A&ARv..18..471A,2016NewAR..72....1G,
  2019A&ARv..27....7C}.  Furthermore, given  the large  densities that
characterize  the  white dwarf  interiors,  these  compact objects  are
considered  reliable cosmic  laboratories to  study the  properties of
baryonic     matter     under      extreme     physical     conditions
\citep{2022FrASS...9....6I}. Among  all the  white dwarfs,  of special
interest  are the  so-called  ultra-massive white  dwarfs, defined  as
those with masses larger than $\sim 1.05 M_\odot$. Ultra-massive white
dwarfs play a key role in constraining the threshold above which stars
explode as supernova to create neutron  stars and they are involved in
extreme  astrophysical  phenomena,  such  as type  Ia  supernovae,  
micronovae explosions, radio     transients     via     an
accretion-induced     collapse \citep{2019MNRAS.490.1166M} as well as
stellar mergers. Ultra-massive  white dwarfs constitute
also powerful tools  to study  the theory  of high  density
plasmas and general relativity.

The theoretical evolution of ultra-massive white dwarfs with masses up
to    $1.29\,   M_\odot$    has    been   studied    in   detail    in
\cite{2019A&A...625A..87C,2022MNRAS.511.5198C}. These studies provide
white dwarf evolutionary sequences  with oxygen-neon (O/Ne) and carbon-oxygen
(C/O) core-chemical  composition,  considering  realistic  initial  chemical
profiles  that  are  the  result  of  the  full  progenitor  evolution
calculated   in  \cite{2010A&A...512A..10S}   and  \cite{ALTUMCO2021},
respectively.
%These  models included  all the relevant  energy sources
%that govern their evolution, such  as the energy released by $^{22}$Ne
%sedimentation  and by  latent heat  and  phase separation  due to  the
%crystallization  process.
This  set   of  ultra-massive  white  dwarf
evolutionary  models provides an  appropriate tool  to study  the
ultra-massive white  dwarf population  in our  Galaxy, subject  to the
condition that white dwarf masses do not exceed $1.29\, M_\odot$.

During the last years, observations of ultra-massive white dwarfs have
been           reported          in           several          studies
\citep{2004ApJ...607..982M,2016IAUFM..29B.493N,2011ApJ...743..138G,2013ApJS..204....5K,2015MNRAS.450.3966B,2016MNRAS.455.3413K,2017MNRAS.468..239C,2021MNRAS.503.5397K,Hollands2020,2021Natur.595...39C,2022MNRAS.511.5462T}. In
particular,    \cite{2018ApJ...861L..13G}    derived   a    mass    of
$1.28\pm0.08\,M_{\odot}$ for  the long  known white  dwarf GD  50. The
number of ultra-massive white dwarfs with mass determinations beyond
$1.29\,  M_\odot$ is  steadily  increasing  with recent  observations.
\cite{2020MNRAS.499L..21P} discovered a rapidly-rotating ultra-massive
white dwarf, WDJ183202.83+085636.24,  with $M=1.33\pm0.01\,M_{\odot}$ meanwhile
\cite{2021Natur.595...39C}    reported    the     existence    of    a
highly-magnetized,  rapidly-rotating  ultra-massive white  dwarf,  ZTF
J190132.9+145808.7,  with   a  mass   of  $\sim   1.327  -   1.365  \,
M_\odot$.  \cite{2021MNRAS.503.5397K} studied  the most  massive white
dwarfs in  the solar  neighborhood and concluded  that other  22 white
dwarfs could  also have masses  larger than $1.29\, M_\odot$,  if they
had    pure    H    envelopes     and    C/O    cores.     Furthermore,
\cite{2022RNAAS...6...36S} has confirmed the  existence of a branch of
faint blue  white dwarfs  in the {\it  Gaia} color  magnitude diagram,
some of them also reported in   \cite{Kilic2020},
which   is  mainly   composed  by
ultra-massive white dwarfs more massive  than $1.29\, M_\odot$.
%Furthermore, \cite{2022RNAAS...6...36S} has reported the existence of a branch of faint blue white dwarfs in the {\it Gaia} color magnitude diagram, which is thought to be composed by ultra-massive white dwarfs with masses larger than $1.29\, M_\odot$. 

%{\bf However, these models are not completely appropriate
%to study the recently detected ultra-massive white dwarfs with stellar masses
%exceeding $1.29\, M_\odot$},
%However, \tiara{there has been indications {\bf (Creo que la palabra adecuada seria "hints". Es un tema delicado. Algunas de estas citas son muy viejas. Segun la fotometria de Gaia, La RE J0317-853 tiene una masa de 1.23 y la PG 1658 + 441 de 1.24, y con barras de error muuuuuuuuy chicas (por eso Kilic no las considera en su muestra). Creo que esta muy bueno poner esto como reseña historica pero hay que ser cuidadosos. Por ahora lo dejo asi y cuando le pegamos una mirada final lo vemos)} in the past of ultra-massive WDs with masses greater than $1.29\, M_\odot$, namely RE J0317-853 \citep{1995MNRAS.277..971B} with a mass of $\sim 1.27-1.4\,M_{\odot}$ determined by \cite{2010A&A...524A..36K} for ONe-core composition, PG 1658 + 441 with $1.31\pm0.02\/M_{\odot}$ \citep{1992ApJ...394..603S} and LHS 4033 with $1.31-1.335\,M_{\odot}$ \citep{2004ApJ...605..400D}. }

In addition  to all  these observations,  gravity($g$)-mode pulsations
have been detected at least in  four ultra-massive white dwarfs 
\citep{1992ApJ...390L..89K,2013ApJ...771L...2H,2017MNRAS.468..239C,2019MNRAS.486.4574R}. Although
these stars have masses slightly  below $1.29~M_\odot$, we expect that
more massive pulsating  white dwarfs will be identified  in the coming
years with the advent of huge volumes of high-quality photometric data
collected by  space missions  such as the  ongoing {\sl  TESS} mission
\citep{2015JATIS...1a4003R}          and         {\sl          Cheops}
\citep{2018A&A...620A.203M} mission, and the  future {\sl Plato} space
telescope \citep{2018EPSC...12..969P}. This  big amount of photometric
data is  expected to make  asteroseismology a promising tool  to study
the structure  and chemical composition of  ultra-massive white dwarfs
\citep{2019A&A...621A.100D,  2019A&A...632A.119C}.  In  fact,  several
successful  asteroseismological analyzes  of  white  dwarfs have  been
carried out  employing data from  space thanks to the  {\sl Kepler/K2}
mission                                    \citep{2010Sci...327..977B,
  2014PASP..126..398H,2020FrASS...7...47C}     and      {\sl     TESS}
\citep{2022arXiv220303769C}.

The increasing number of detected ultra-massive white dwarfs with masses beyond
$1.29\,M_{\odot}$  as  well as  the  immediate  prospect of  detecting
pulsating  white  dwarfs  with  such masses,  demand  new  appropriate
theoretical   evolutionary   models   to   analyze   them.   Recently,
\cite{2021ApJ...916..119S} has  studied the evolution of  white dwarfs
more massive than $1.29\, M_\odot$  with the focus on neutrino cooling
via the Urca  process, showing that this process is  important for age
determination  of  O/Ne-core  white  dwarf  stars.  These  models  were
calculated  employing  the set  of  standard  equations to  solve  the
stellar  structure and  evolution  under the  assumption of  Newtonian
gravity.  However,  the  importance  of  general  relativity  for  the
structure  of  the most  massive  white  dwarfs cannot  be  completely
disregarded. This was recently assessed by \cite{2018GReGr..50...38C},
who solved  the general relativistic hydrostatic  equilibrium equation
for a completely degenerate ideal Fermi electron gas. They demonstrate
that for fixed  values of total mass, large deviations  (up to 50$\%$)
in the Newtonian white dwarf radius are expected, as compared with the
general relativistic  white dwarf  radius. The  impact of  a non-ideal
treatment  of  the  electron  gas  on  the  equilibrium  structure  of
relativistic  white dwarfs  was studied  by \cite{2011PhRvD..84h4007R}
and \cite{2017RAA....17...61M}, who  derived the mass-radius relations
and  critical masses  in the  general relativity  framework for  white
dwarfs of different core chemical compositions. These studies conclude
that general  relativistic effects are relevant  for the determination
of the radius of massive white dwarfs. \cite{2014PhRvC..89a5801D} and,
more  recently,   \cite{2021ApJ...921..138N}  have   investigated  the
general  relativity  effects  in  static  white  dwarf  structures  of
non-ideal   matter  in   the   case  of   finite  temperature.   While
\cite{2014PhRvC..89a5801D} focused their work on the effects of finite
temperature      on     extremely      low-mass     white      dwarfs,
\cite{2021ApJ...921..138N} studied the stability  of massive hot white
dwarfs   against  radial   oscillations,  inverse   $\beta-$decay  and
pycnonulcear reactions. They  find that the effect  of the temperature
is still  important for determining  the radius of very  massive white
dwarfs.

Despite several works have been devoted to the study of the effects of
general relativity  on the  structure of white  dwarfs, none  of these
works has  calculated the evolution  of such structures.  Moreover, in
all of the  works mentioned above, the white dwarf  models are assumed
to  be composed  by solely  one chemical  element. The  exact chemical
composition determines  both the  mass limit of  white dwarfs  and the
nature of  the instability  (due to  general-relativity effects  or to
$\beta$-decays, e.g. \citealt{2011PhRvD..84h4007R}).  In this paper we
compute the first  set of constant rest-mass  ultra-massive O/Ne white
dwarf evolutionary models which fully take into account the effects of
general    relativity   on    their   structural    and   evolutionary
properties.  Furthermore,  we   consider  realistic  initial  chemical
profiles  as  predicted by  the  progenitor  evolutionary history.  We
employ the La  Plata stellar evolution code, {\tt  LPCODE}, to compute
the full  evolutionary sequence of  1.29, 1.31, 1.33, 1.35,  and 1.369
$M_{\sun}$ white  dwarfs. The standard equations  of stellar structure
and evolution  solved in this code  have been modified to  include the
effects  of  general relativity.  For  comparison  purposes, the  same
sequences have  been computed but  for the Newtonian gravity  case.
%We
%have included the energy  released during the crystallization process,
%both   due  to   latent  heat   and  due   to  the   induced  chemical
%redistribution.  This  chemical   redistribution  is  responsible  for
%reducing  the  stellar  radius,   thus  being  crucial  to  accurately
%determine the  mass-radius relation for these  objects \citep[see][for
%details]{2019A&A...625A..87C}.
  We  assess  the  resulting  cooling
times  and provide  precise time  dependent mass-radius  relations for
relativistic ultra-massive  white dwarfs.  We also  provide magnitudes
in Gaia, Sloan Digital Sky  Survey and Pan-STARRS passbands, using the
model  atmospheres of  \cite{2010MmSAI..81..921K,2019A&A...628A.102K}.
This set of cooling sequences,  together with the models calculated in
\cite{2019A&A...625A..87C}  and \cite{2022MNRAS.511.5198C},  provide a
solid theoretical framework to study  the most massive white dwarfs in
our Galaxy.

This paper is organized as follows. In Sect. \ref{equations}
we describe the modifications to our code to incorporate the effects of general relativity. 
In Sect. \ref{models} we detail the main constitutive physics
of our white dwarf sequences. Sect. \ref{results} is devoted to describe the impact of general relativity effects on the relevant evolutionary properties of our massive white dwarfs. In this section we also compare and discuss the predictions of our new white dwarf sequences with  observational data of ultra-massive white dwarfs, in particular with the recently reported faint blue branch of ultra-cool and ultra-massive objects revealed by {\it Gaia} space mission.
Finally, in Sect. \ref{conclusions} we summarize the main finding of the paper.

\section{The equations of stellar structure and evolution in general relativity}
\label{equations}

Our set of ultra-massive O/Ne white dwarf evolutionary sequences  has been computed with 
the stellar evolution code  {\tt LPCODE} developed by La Plata group, which  has  been widely used 
and tested in numerous  stellar evolution contexts of low-mass stars and particularly  in white dwarf stars \citep[see][for         details]
{2003A&A...404..593A,2005A&A...435..631A, 2013A&A...555A..96S, 2015A&A...576A...9A,
2016A&A...588A..25M,2020A&A...635A.164S,
2020A&A...635A.165C}.  For this work, the stellar structure and evolution equations have 
been modified to include the effects of general relativity, following the formalism given in \cite{1977ApJ...212..825T}. Within this formalism, the fully general relativistic partial differential equations governing the evolution of a spherically symmetric star are presented in a way they resemble the standard Newtonian equations of stellar structure \citep{2012sse..book.....K}. Specifically, the structure and evolution of the star is specified by the Tolman-Oppenheimer-Volkoff (TOV) equation of hydrostatic equilibrium, the equation of mass distribution, the luminosity equation, and the energy transport equation:

\begin{equation}
\frac{\partial P}{\partial m}= -\frac{G m}{4 \pi r^4}\ \mathscr{H G V} \ ,
\label{TOV}
\end{equation}

\begin{equation}
\frac{\partial r}{\partial m}= (4 \pi r^2 \varrho\ \mathscr{V})^{-1} \ ,
\label{MD}
\end{equation}

\begin{equation}
\frac{1}{\mathscr{R}^2}\frac{\partial (L \mathscr{R}^2)}{\partial m}=  -\varepsilon_\nu - \frac{1}{\mathscr{R}} 
{\frac{\partial u} {\partial t}} + \frac{1}{\mathscr{R}} {\frac{P} {\varrho^2}\frac{\partial \varrho} {\partial t}}\ ,
\label{lumistandard}
\end{equation}

\begin{equation}
\frac{\partial (T \mathscr{R})}{\partial m}= -\frac{3}{64 \pi^2 ac}\frac{\kappa L}{r^4 T^3}\mathscr{R} \qquad {\rm if} \quad \nabla_{\rm rad} \leq \nabla_{\rm ad} \ ,
\label{T_rad}
\end{equation}

\begin{equation}
\frac{\partial \rm{ln} T}{\partial m}= \nabla\ \frac{\partial \rm {ln} P}{\partial m} \qquad {\rm if} \quad \nabla_{\rm rad} > \nabla_{\rm ad} \ ,
\label{T_conv}
\end{equation}
where $t$ is the Schwarzschild time coordinate, $m$ is the rest mass inside a radius $r$ or baryonic mass, i.e., the mass of one hydrogen atom in its ground state
multiplied by the
total number of baryons inside $r$,  and $\varrho$ is the density of rest mass. During the entire cooling process, the 
total baryonic mass remains constant. $c$ is the speed of light, $u$ is the internal energy per unit mass, and $\varepsilon_\nu$ is the energy lost by neutrino emission per unit mass. $\mathscr{H, G, V,}$ and $\mathscr{R}$ are the dimensionless general relativistic correction factors, which turn to unity in the Newtonian limit. These factors correspond, respectively, to
the enthalpy, gravitational acceleration, volume, and redshift correction factors, and  are  given by

\begin{align}
\mathscr{H} &= \frac{\varrho^t}{\varrho} + \frac{P}{\varrho c^2},\\
\mathscr{G} &= \frac{ m^t + 4 \pi r^3 P/c^2} {m},   \\
\mathscr{V} &= \left(1 - \frac{ 2 G m^t}{ r c^2}\right)^{-1/2}, \\
\mathscr{R} &= e^{\Phi/c^2}, \\
\label{R-fact}
\end{align}

\noindent where  $m^t$ is the mass-energy inside a radius $r$ and includes contributions from the
rest-mass energy, the internal energy, and the gravitational  potential energy, which is negative.  $\varrho^t$ 
is the density of total nongravitational mass-energy, and includes the density of rest mass plus contributions from
kinetic and potential energy density due to particle interactions (it does not include the gravitational potential energy density), that is $\varrho^t= \varrho + (u \varrho)/ c^2 $. 
Since the internal and
gravitational potential energy change during the course of evolution, the stellar
mass-energy is not a conserved quantity.  $\Phi$ is the general 
relativistic gravitational potential related to the temporal metric coefficient. At variance with
the Newtonian case, the gravitational potential
appears explicitly in the evolution equations. We note that 
the TOV hydrostatic equilibrium equation differs
markedly from its Newtonian counterpart, providing a steeper
pressure gradient. Also we note that the presence of  $\mathscr{V}$ in that equation prevents  $m^t$ from being larger than $rc^2/2G$. 

The radiative gradient $\nabla_{\rm rad}$ is given by 

\begin{equation}
\nabla_{\rm rad} = \frac{3}{16 \pi ac G}\frac{\kappa L P}{m T^4}\frac{1}{\mathscr{H G V}} + \left(1 - \frac{\varrho^t/\varrho}
{\mathscr{H}} \right) \ .
\end{equation}

In Eq. (\ref{T_conv}),  $\nabla$ is the convective temperature gradient, which, in the present work, is  given by the solution of the mixing length
theory. We mention that in ultra-massive white dwarfs the occurrence of convection is restricted exclusively to a very narrow
outer layer\footnote{This may not be true if neutrino cooling via the
  Urca process is considered, in which case an inner convection zone is expected, see \cite{2021ApJ...916..119S}.}, being mostly adiabatic. We follow \cite{1977ApJ...212..825T} to generalize the mixing length
theory to general relativity.  In Eq. (\ref{lumistandard}) we have omitted the energy generation by nuclear reactions since these are not happening in our models. However, they should be added when taking into account Urca processes.% and inverse beta decays?

To solve Eqs. (\ref{TOV})-(\ref{T_conv}) we need two additional equations that relate   $m^t$  and  $\Phi$ with  $m$. These two equations, which
are not required in the Newtonian case,  have to be solved simultaneously with Eqs. (\ref{TOV})-(\ref{T_conv}). These extra equations are given
by  \citep[see][]{1977ApJ...212..825T}

\begin{equation}
\frac{\partial m^t }{\partial m}= \frac{\varrho^t}{\varrho} \frac{1}{\mathscr{V}}\ ,
\label{grav_mass}
\end{equation}

\begin{equation}
\frac{\partial \Phi}{\partial m}= \frac{G m}{4 \pi r^4 \varrho}\ \mathscr{G V} \ .
\label{phi}
\end{equation}

\subsection{Boundary conditions}

The rest mass, total mass-energy, and radius of the star correspond, respectively, to the values of $m$, $m^t$, and $r$ at the surface of the star. We  denote them by

\begin{equation}
M_{\rm WD}=m \ ,  \qquad  M_{\rm G}=m^t \ ,  \qquad R=r \qquad {\rm at \ the \ surface} \ .
\end{equation}

$M_G$ is the total gravitational mass, i.e., the stellar mass that would be measured by a distant observer, which turns out to be
less than the total baryonic mass of the white dwarf. 
Outer boundary conditions for our evolving models are provided
by the integration of

\begin{equation}
\frac{d P}{d \tau}= \frac{g^t}{\kappa} \ ,
\label{atm}
\end{equation}

\noindent and assuming a gray model atmosphere. $\tau$ is the optical depth and $g^t$ is the "proper" surface gravity of the star (as measured on the
surface) corrected by general relativistic effects and given by

\begin{equation}
g^t= \frac{G M_G}{R^2} \mathscr{V} \ .
\label{grav}
\end{equation}

\begin{figure*}
        \centering
        \includegraphics[width=1.5\columnwidth]{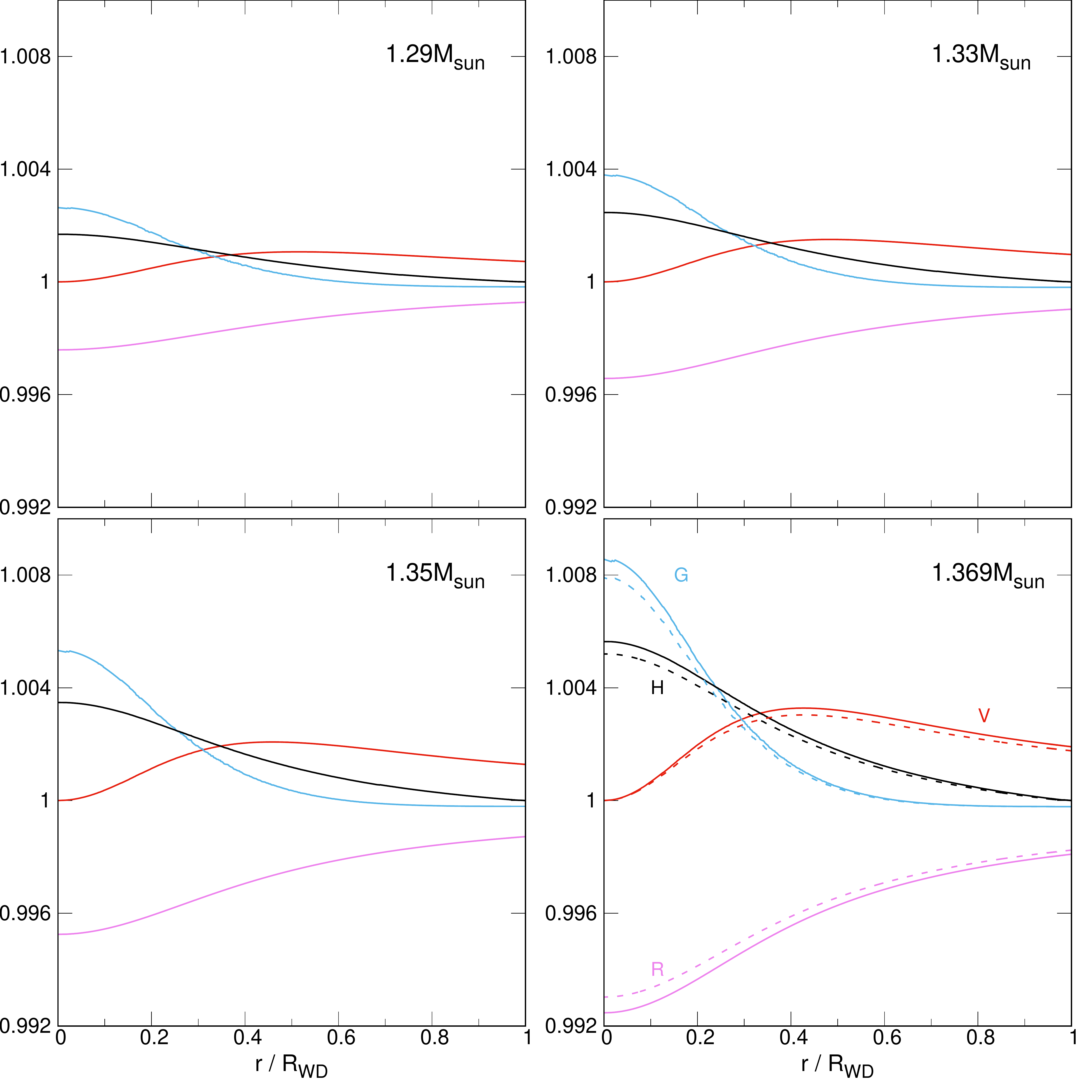}
        \caption{Run of general relativistic correction factors $\mathscr{H, G, V}$, and $\mathscr{R}$ (black, blue, red, and pink lines, respectively)
        for 1.29, 1.33, 1.35, and 1.369$M_{\sun}$ white dwarf
        models at log $L/L_{\sun}=-3$ in terms of the fractional radius. 
        Dashed lines in the bottom right panel illustrate the behavior of the same factors         for a 1.369$M_{\sun}$ model at log $L/L_{\sun}=-0.4$.} 
        \label{factors}
\end{figure*}

In addition, the general relativistic metric for spacetime in the star interior must match to the metric outside created by the star (Schwarzschild
metric). The match requires  that $\Phi$ satisfies the surface boundary condition 

\begin{equation}
\Phi= \frac{1}{2} c^2 \ln \left(1-\frac{2 G M_G }{R c^2} \right)\qquad {\rm at} \quad m=M_{\rm WD} \ .
\label{phi_sup}
\end{equation}

At the stellar center, $m=0$, we have $m^t=0$, $r=0$, and $L=0$.

\section{Initial models and input physics}
\label{models}

We have  computed the full  evolution of  1.29, 1.31, 1.33,  1.35, and
1.369 $M_{\sun}$  white dwarfs assuming  the same O/Ne  core abundance
distribution for  all of them. The adopted  core composition
corresponds to that  of the 1.29 $M_{\sun}$  hydrogen-rich white dwarf
sequence  considered  in  \cite{2019A&A...625A..87C}, which  has  been
derived from the evolutionary history  of a 10.5 $M_{\sun}$ progenitor
star \citep{2010A&A...512A..10S}.  In  this  work,  we
restrict ourselves  to O/Ne-core massive white  dwarfs, thus extending
the  range  of   O/Ne  white  dwarf  sequences   already  computed  in
\cite{2019A&A...625A..87C} in the frame of Newtonian theory of stellar
interior.  O/Ne  core  white  dwarfs  are  expected  as  a  result  of
semi-degenerate  carbon   burning  during  the  single   evolution  of
progenitor  stars that  evolve to  the Super  Asymptotic Giant  Branch
\citep{1997ApJ...485..765G,2005A&A...433.1037G,2006A&A...448..717S,2010MNRAS.401.1453D,2011MNRAS.410.2760V}.
Recent calculations of the remnant of a double white dwarf merger also
predict O/Ne core composition as a result of off-center carbon burning
in  the merged  remnant, when  the remnant  mass is  larger than  1.05
M$_\sun$  \citep[see][]{2021ApJ...906...53S}.  In  particular,  it  is
thought  that  a considerable  fraction  of  the massive  white  dwarf
population   is    formed   as    a   result   of    stellar   mergers
\citep{2020A&A...636A..31T,2020ApJ...891..160C,2022MNRAS.511.5462T}.
We note however that the  existence of ultra-massive white dwarfs with
C/O  cores   resulting  from  single  evolution   cannot  be
discarded \cite[see][]{2021A&A...646A..30A, 2022arXiv220202040W}.
%{\bf ACA CITAR CAMISASSA 2022b, si es que nos llega el referato.}

The adopted input physics for our relativistic white dwarf models is the
same as that in \cite{2019A&A...625A..87C}.  In brief, the equation of
state     for     the     low-density    regime     is     that     of
\cite{1979A&A....72..134M}, and that of \cite{1994ApJ...434..641S} for
the high-density  regime, which takes  into account all  the important
contributions  for  both  the  solid and  liquid  phases.  We  include
neutrino emission for pair,  photo, and Bremsstrahlung processes using
the rates of \cite{1996ApJS..102..411I}, and of \cite{1994ApJ...425..222H}
for plasma processes.    The  energetics
resulting from  crystallization processes  in the core  has been
included  as in  \cite{2019A&A...625A..87C}, and  it is  based on  the
two-component  phase diagram  of dense  O/Ne mixtures  appropriate for
massive white  dwarf interiors, \cite{2010PhRvE..81c6107M}. As shown
by \cite{2021ApJ...919...87B}, $^{23}$Na and $^{24}$Mg impurities have
only  a  negligible   impact  on  the  O/Ne  phase   diagram  and  the
two-component  O/Ne phase  diagram can  be safely  used to  assess the
energetics resulting from  crystallization.  We have not
considered  the energy  released by  $^{22}$Ne sedimentation  process,
since it is negligible in O/Ne white dwarfs \citep{2021A&A...649L...7C}.

\begin{figure}
        \centering
        \includegraphics[width=1.\columnwidth]{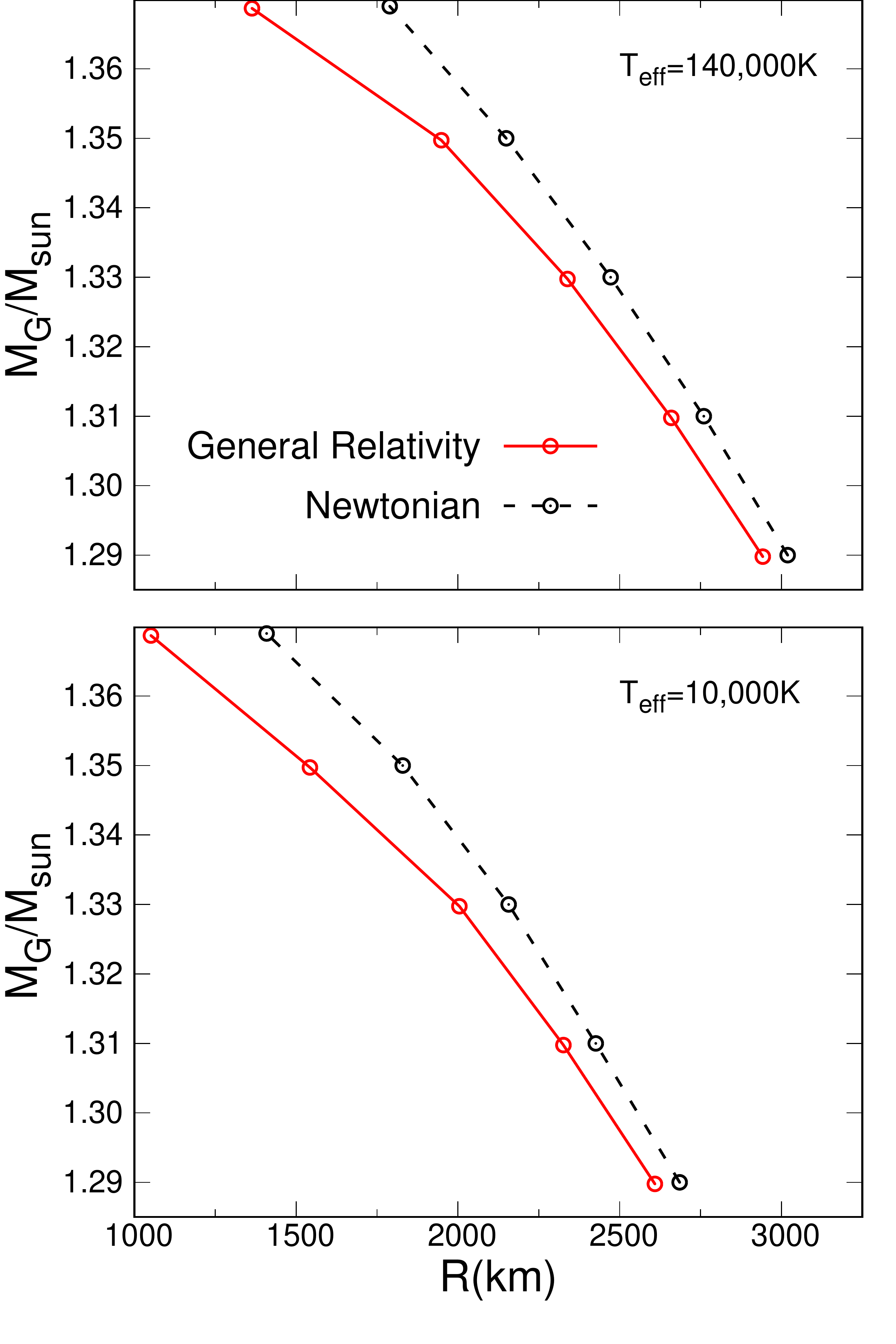}
        \caption{The gravitational mass versus the stellar radius for our O/Ne ultra-massive white dwarf models considering (red symbols and lines) and disregarding (black symbols and lines) the effects of general relativity at two different effective temperatures.  
        } 
        \label{mr}
\end{figure}

\section{General relativity effects on the evolution of massive white dwarfs}
\label{results} 

\begin{table*}[t]
\centering
\begin{tabular}{lccccccc} 
  \hline
 \hline  \\[-4pt]           
$M_{\rm WD}$ & $M_{\rm G}$  & $R^{\rm Newt}$ & $R^{\rm GR}$  & log g$^{\rm Newt}$ &  log g$^{\rm GR}$ & $\varrho_c^{\rm Newt}$  & $\varrho_c^{\rm GR}$\\
  $M_\odot$ &   $M_\odot$ & km & km &  cm s$^{-2}$  &    cm s$^{-2}$   & g cm$^{-3}$   & g cm$^{-3}$  \\
  \hline \\[-4pt]  
  1.29   & 1.28977 &  2685.40  &  2608.86  &   9.375     &   9.401 &  $ 6.71\times 10^{8}$  &                   $7.51 \times 10^{8}$     \\
  1.31   & 1.30976 &  2426.04 &  2326.17  &   9.470    &   9.507  &   $ 9.98 \times 10^{8}$  &                   $1.17 \times 10^{9}$     \\
  1.33   & 1.32974 &  2156.90 &  2004.60   &   9.579    &   9.643  &   $ 1.57 \times 10^{9}$  &                  $2.06 \times 10^{9}$    \\
  1.35   & 1.34972 &  1829.29 &  1542.51  &   9.728    &   9.878  &   $ 2.90 \times 10^{9}$ &   $5.36 \times 10^{9}$     \\
  1.369  & 1.36871 &  1408.77 &  1051.16  &   9.961    &  10.217  &    $7.42 \times 10^{9}$  &      $2.11 \times 10^{10}$     \\
  \hline  
  \\
  \end{tabular}                 
  \caption{Relevant characteristics of our sequences a $T_{\rm eff}$=10,000K.  $M_{\rm WD}$: total baryonic
  mass.  $M_{\rm G}$: total gravitational mass. $R^{\rm Newt}$: stellar radius in the Newtonian case. $R^{\rm GR}$: 
  stellar radius in the general relativity case. g$^{\rm Newt}$: surface gravity in the Newtonian case. g$^{\rm GR}$: surface
  gravity in the general relativity case. $\varrho_c^{\rm Newt}$: central density of rest mass in the Newtonian case.  $\varrho_c^{\rm GR}$:
  central density of rest mass in the general relativity case. }
    \label{table1}
\end{table*}

Here,  we describe  the impact  of general  relativity effects  on the
relevant properties of our  constant rest-mass evolutionary tracks. We
begin  by examining  Fig.  \ref{factors}, which  displays the  general
relativistic correction factors $\mathscr{H, G, V}$, and $\mathscr{R}$
(black,  blue, red,  and pink  lines,  respectively) in  terms of  the
fractional radius for the 1.29,  1.33, 1.35, and 1.369$M_{\sun}$ white
dwarf    models    at  log $L/L_{\sun}=-3$. Dashed  lines in  the
bottom right  panel illustrate
the run of the same factors for a 1.369$M_{\sun}$ white dwarf model at
log $L/L_{\sun}=-0.4$ (log $T_{\rm  eff}=5$). 
We  recall that these
factors are unity in the  Newtonian limit. As expected, the importance
of  general relativistic  effects  increases as  the  stellar mass  is
increased.  We note  that $\mathscr{V}$  is  unity at  the center  and
attains  a  maximum  value  at  some inner  point  in  the  star.  The
relativistic  factor   $\mathscr{R}$  decreases  towards   the  center,
departing  even   more  from  unity,  meanwhile   the  other  factors,
$\mathscr{G}$  and $\mathscr{H}$  increase towards  the center  of the
star. The behavior  of the  relativistic correction  factors can  be
  traced back  to curvature effects, as well as  the  fact that  the
  pressure and the internal
energy appear  as a  source for
gravity   in   general   relativity.   For   maintaining   hydrostatic
equilibrium,  then, both  density and  pressure gradients  are steeper
than in  Newtonian gravity. This  makes the factors  $\mathscr{G}$ and
$\mathscr{H}$,  which  depend directly  on  density  and pressure,  to
increase  towards the  center  of the  star.  The relativistic  factor
$\mathscr{V}$, which can be interpreted as a correction to the volume,
would be unity at the center of the star where the volume is zero, and
increase because  of the  increasing of  density in  general relativity
respect  to  the  density  in   Newtonian  gravity.  However,  as  the
departures from the Newtonian case decrease towards the surface of the
star, $\mathscr{V}$  decreases  towards the outside,  achieving a
maximum  value in  between.   We note  that  the relativistic  factors
depend slightly on the effective temperature.
 
The impact of relativistic effects  on the mass-radius relation
at two different effective temperatures
can  be  appreciated in  Fig.\,\ref{mr}.  We  note  that for
the most massive white dwarfs, at  a  given
gravitational mass,  the radius is  markedly smaller in the  case that
the general relativity  effects are taken into  account.
At   a  stellar  mass  of  1.369 $M_{\sun}$ the
stellar radius  becomes only 1050  km, 25\%  smaller than predicted  by the
Newtonian  treatment (see  Table  \ref{table1}). As  in the  Newtonian
case, the effect of finite temperature  on the stellar radius is still
relevant in very  massive white dwarfs.  We mention that general relativistic
  corrections become negligible for stellar masses smaller than
  $\approx$ 1.29 $M_{\sun}$. In particular, for stellar masses below that value,
  the stellar radius  results below  2
  \%  smaller
when general relativity effects are taken into account . 

In  our calculations,  O/Ne white
dwarfs  more  massive  than 1.369  $M_{\sun}$  become  gravitationally
unstable (which occurs at a given finite central density) with respect
to  general relativity  effects, in  agreement with  the findings  for
zero-temperature models  reported in \cite{2011PhRvD..84h4007R}  for a
pure-oxygen     white      dwarf     (1.38024      $M_{\sun}$)     and
\cite{2017RAA....17...61M} for white dwarfs composed of oxygen (1.3849
$M_{\sun}$) or of neon (1.3788  $M_{\sun}$), although their values are
slightly higher\footnote{Preliminary computations we  performed for
  oxygen-rich core white dwarfs show that they become
  unstable  at  1.382  $M_{\sun}$.}. We mention that  for the 1.369
$M_{\sun}$  white dwarf  model,  the  central  density in  the
general relativity case reaches $2.11 \times 10^{10}$ g cm$^{-3}$ (see
Table \ref{table1}).  Such  density is near the  density threshold for
inverse $\beta-$decays.  We have not considered that matter inside our
white  dwarf models  may  experience instability  against the  inverse
$\beta-$decay.  O or  Ne white dwarfs are expected  to become unstable
against the inverse  $\beta-$decay process at a stellar  mass near the
critical mass resulting from general  relativity effects, of the order
of       1.37      $M_{\sun}$       \citep[see][]{2011PhRvD..84h4007R,
  2017RAA....17...61M}.

\begin{figure}
        \centering
        \includegraphics[width=1.\columnwidth]{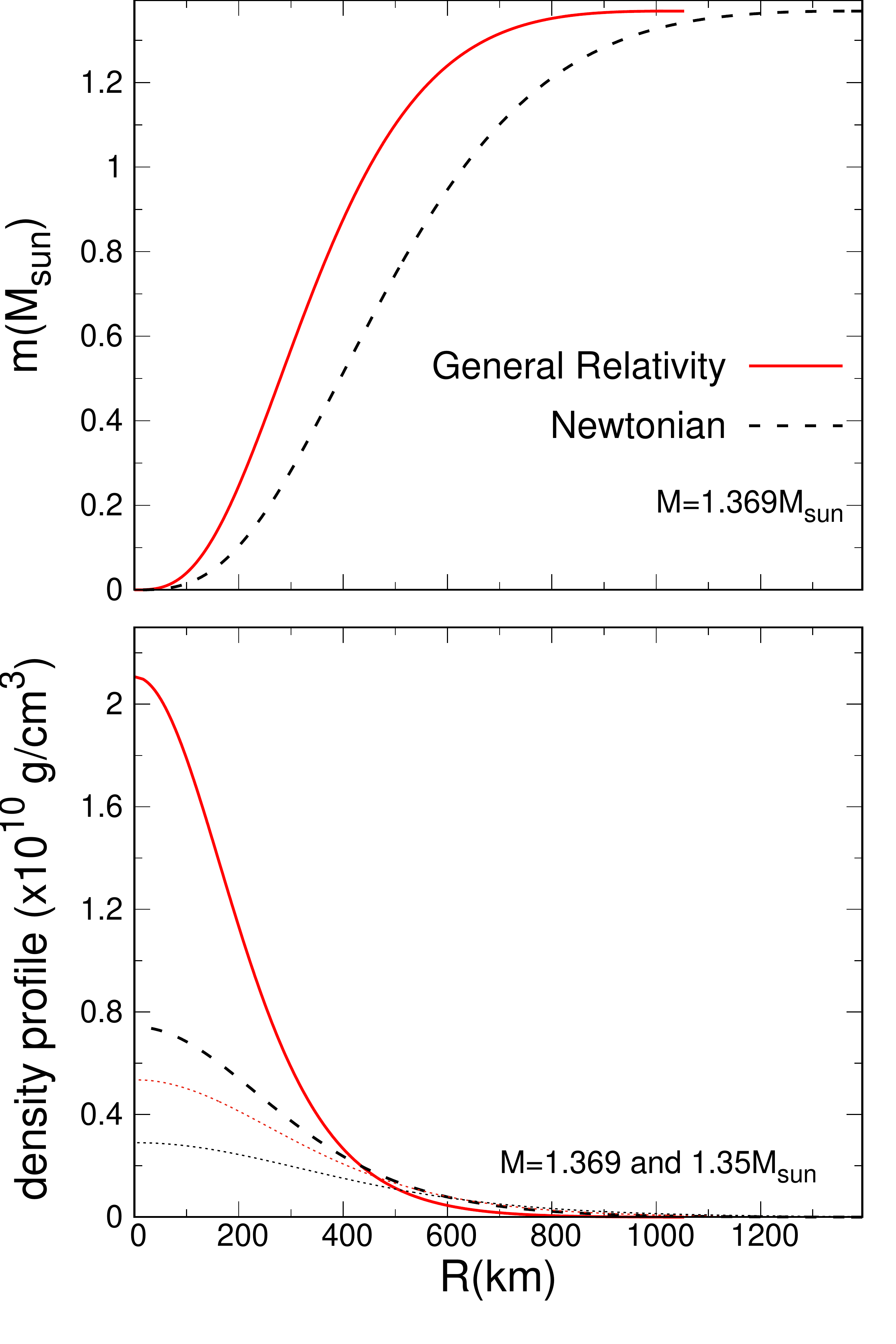}
        \caption{Rest mass $m$ (upper panel) and density of rest mass (bottom panel)  for
         the general relativity and Newtonian cases (red solid and black dashed lines, respectively) in terms of radial coordinate 
         for 1.369$M_{\sun}$  white dwarf models at advanced stage of evolution. Dotted lines
         in the bottom panel depict the situation for the 1.35$M_{\sun}$  models.} 
        \label{mass-density}
\end{figure}

\begin{figure}
        \centering \includegraphics[width=1.\columnwidth]{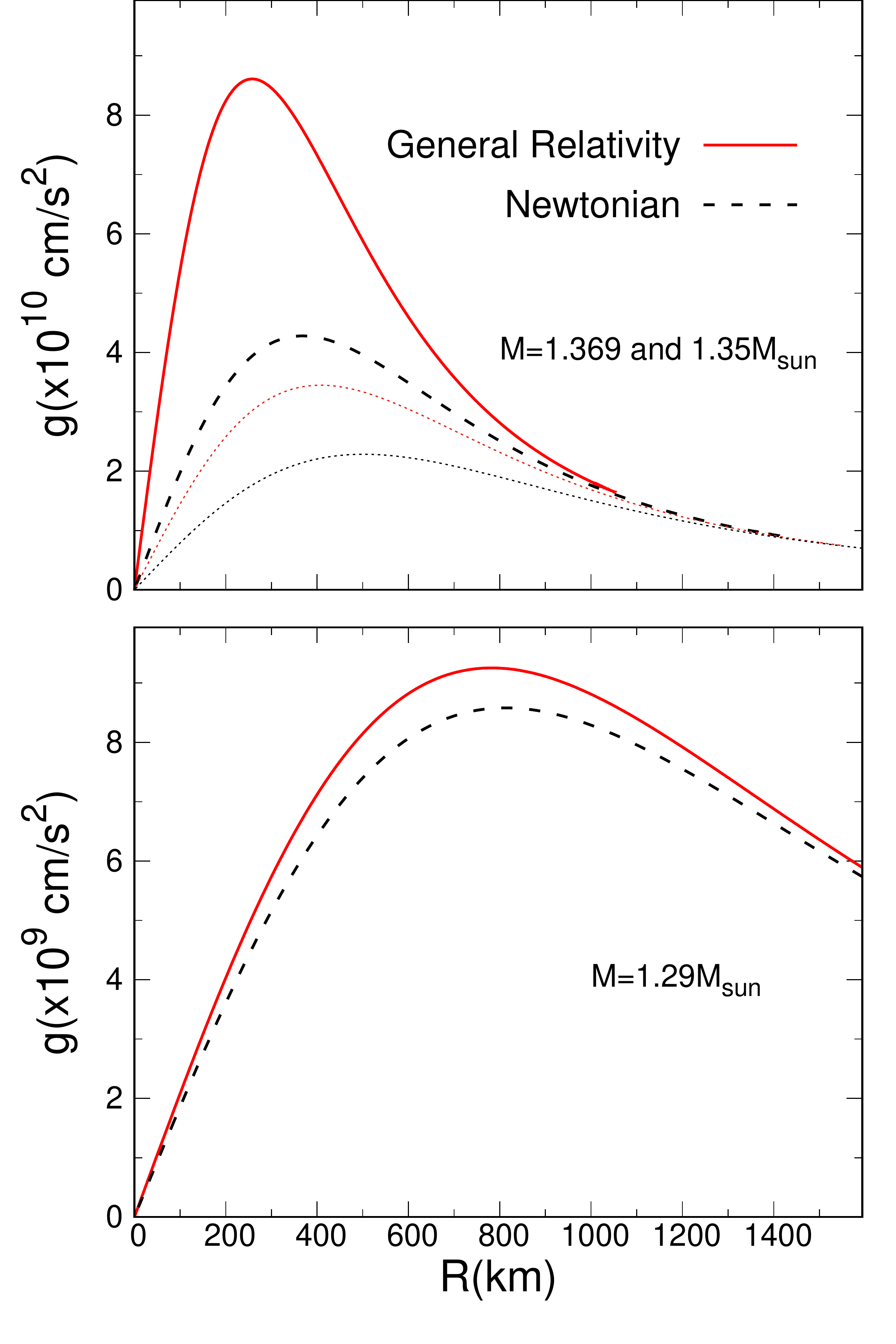}
        \caption{General relativity and  Newtonian gravitational field
          (red  solid and  black  dashed  lines, respectively)  versus
          radial  coordinate  for 1.369$M_{\sun}$  and  1.29$M_{\sun}$
          white dwarf  models at  advanced stage of  evolution. Dotted
          lines  in  the upper  panel  depict  the situation  for  the
          1.35$M_{\sun}$ models.}
        \label{gravity}
\end{figure}

The  inner profile  of rest  mass  and density  of rest  mass for  the
1.369$M_{\sun}$  white  dwarf  model  in the  general  relativity  and
Newtonian  cases  are   shown  in  the  upper  and   bottom  panel  of
Fig. \ref{mass-density},  respectively.  For such massive  white dwarf
model,  general   relativity  effects   strongly  alter   the  stellar
structure,  causing matter  to be  much more  concentrated toward  the
center of the  star and the central  density to be larger  than in the
Newtonian case.  The impact  remains noticeable towards  lower stellar
masses, although to a  lesser extent, as can be noted  for the case of
1.35$M_{\sun}$  white  dwarf  model  shown  in  the  bottom  panel  of
Fig. \ref{mass-density} (dotted  lines).  In view of this,  the run of
the  gravitational field  versus radial  coordinate for  the general relativity
case  differs markedly from that  resulting from the Newtonian
case.  This  is  shown  in  Fig.  \ref{gravity}  for  1.369$M_{\sun}$,
1.35$M_{\sun}$, and 1.29$M_{\sun}$ white  dwarf models. In particular,
the gravitational field  in the general relativistic  case as measured
far from the star is given by

\begin{equation}
g^{\rm GR}= \frac{G m}{r^2} \mathscr{G} \mathscr{V}^2 \ .
\label{grav_sup}
\end{equation}

Clearly, the gravitational field in the  most massive of our models is
strongly affected  by general  relativity.  In the  stellar interior,
large  differences  arise  in  the gravitational  field  due  to  the
inclusion of general relativity effects.  We note that such differences
do not  arise from the relativistic  correction factors $\mathscr{G}
\mathscr{V}^2   $  (see   Fig.    \ref{factors})   to  the   Newtonian
gravitational field $g^{\rm New}= G  m /r^2$ that appear explicitly in
Eq.  (\ref{grav_sup}),  but  from  the solution  of  the  relativistic
equilibrium instead, which  gives a different run  for $m(r)$ compared
to the Newtonian case.

\begin{figure}
        \centering
        \includegraphics[width=1.\columnwidth]{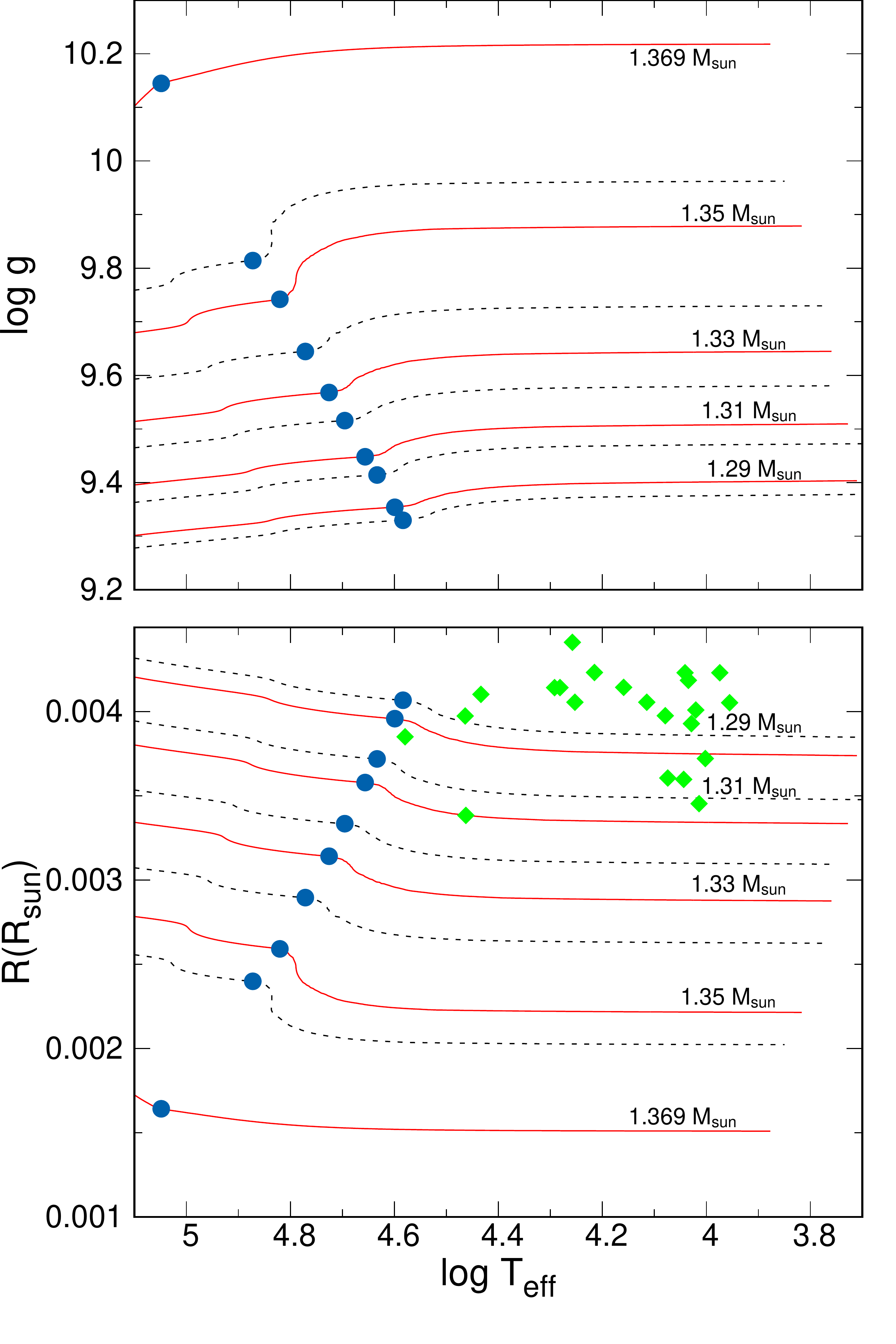}
        \caption{Surface gravity and stellar radius (in solar units) in terms
        of the effective temperature for all of our
        sequences are displayed in the upper and bottom panels, respectively. Red solid and black dotted lines correspond to the general relativity  and Newtonian cases, respectively. From bottom (top) to top (bottom), curves in the upper (bottom) panel  correspond to
 1.29, 1.31, 1.33, 1.35, and 1.369$M_{\sun}$   white dwarfs cooling sequences. Blue filled circles denote the onset of core crystallization in each sequence.  The most massive white dwarfs in the solar neighborhood analyzed in \cite{2021MNRAS.503.5397K} are displayed using green filled diamonds.}  
        \label{grav-surface}
\end{figure}

Additionally,  the surface  gravity  and stellar  radius are  
affected by  the effects of  general relativity. These  quantities are
shown in Fig. \ref{grav-surface} in terms of the effective temperature
for  all of  our sequences  for the  general relativity  and Newtonian
cases, using solid and dashed lines, respectively. In the most massive
sequences,  general  relativity  effects markedly  alter  the  surface
gravity  and   stellar  radius. In this sense,  we  infer that
general relativity
effects lead to a stellar mass value about 0.015$M_{\sun}$  smaller
for  cool white dwarfs with  measured surface gravities of log g $\approx$ 10.
The  photometric   measurements  of
\cite{2021MNRAS.503.5397K} for  the radius of the  ultra-massive white
dwarfs in the solar neighborhood are  also plotted in this figure. For
the  more massive  of such  white dwarfs,  the stellar  radius results
2.8-4$\%$  smaller  when general  relativity  effects  are taken  into
account.

We note that most of our  sequences display a sudden increase in their
surface  gravity   at  high   effective  temperatures.  As   noted  in
\cite{2019A&A...625A..87C},  this  is related  to  the  onset of  core
crystallization  (marked with  blue  filled circles  in each  sequence
depicted in Fig. \ref{grav-surface}),  which modifies the distribution
of $^{16}$O  and $^{20}$Ne.  Specifically, the abundance  of $^{20}$Ne
increases in the core of  the white dwarf as crystallization proceeds,
leading to larger Coulomb interactions and hence to denser cores, and,
therefore,  to higher  surface gravities.  This behavior  can also  be
regarded   as   a   sudden    radius   decrease   (bottom   panel   of
Fig. \ref{grav-surface}).  In this context,  we note that  the density
increase due to the increase in the core abundance of $^{20}$Ne during
crystallization eventually causes O/Ne white dwarf models with stellar
masses larger  than $\gtrsim 1.36  M_\sun $ to  become gravitationally
unstable against general relativity effects.   In order to explore the
mass range  of stable white dwarfs  in the absence of  this processes,
the  1.369$M_{\sun}$ relativistic  sequence was  computed disregarding
the  effect  of   phase  separation  (but  not   latent  heat)  during
crystallization.

\subsection{General relativity effects on the white dwarf cooling times}

\begin{figure}
        \centering
        \includegraphics[width=1.\columnwidth]{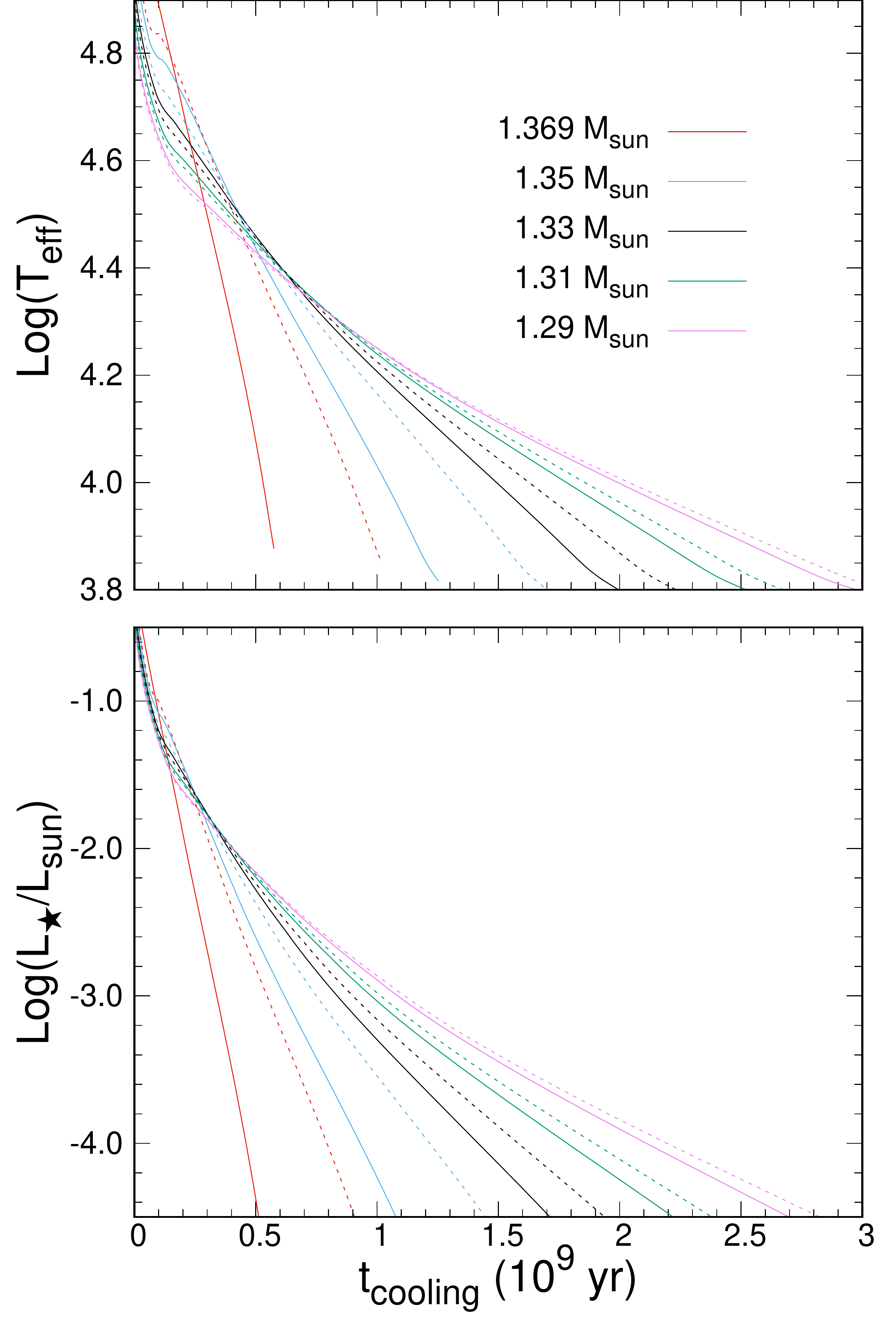}
        \caption{Effective temperature and surface luminosity (upper and bottom panels) versus the cooling times for our  1.29, 1.31, 1.33, 1.35, and 1.369$M_{\sun}$   white dwarfs sequences.  
        Solid (dashed) lines correspond to the general relativity (Newtonian) cases.
        Cooling time is counted from the time of white dwarf formation.} 
        \label{age}
\end{figure}

The  cooling properties  of the  ultra-massive white  dwarfs are  also
markedly altered by  general relativity effects, in  particular the
 the most massive ones. This is illustrated in Fig. \ref{age},
which  compares  the  cooling  times  of our  models  for  the  general
relativity   and    Newtonian   cases,   solid   and    dashed   lines
respectively. The  cooling times are set  to zero at the  beginning of
cooling  tracks  at  very high  effective  temperatures.  Gravothermal
energy is the  main energy source of the white  dwarfs, except at very
high  effective   temperatures  where   energy  released   during  the
crystallization  process  contributes  %substantially  to  the  energy
to the budget     of    the     star.     As      noticed     in
\cite{2021A&A...649L...7C}, ultra-massive  O/Ne-core white  dwarfs
evolve   significantly  fast  into  faint
magnitudes.    General   relativity  effects   cause
ultra-massive white dwarfs to evolve  faster than in the Newtonian
case  at advanced  stages  of evolution.   In  particular, the  $1.369
M_\sun$  relativistic sequence  reaches $\log(L/L_\sun)$=-4.5  in only
$\sim 0.5$  Gyrs, in contrast with  the $\sim 0.9$ Gyrs  needed
in the Newtonian  case. The  larger  internal
densities inflicted by general relativity make the Debye cooling phase
more relevant than in the Newtonian case at a given stellar mass, thus
resulting in a  faster cooling for the sequences  that include general
relativity effects. 
The fast cooling  of these
objects, together with their low  luminosity and rare formation rates,
would  make  them  hard to  observe.  The  trend in
the cooling behavior  is reversed at
earlier  stages  of evolution,  where  white  dwarfs computed  in  the
general   relativity  case   evolve   slower   than  their   Newtonian
counterparts. This  is because  white dwarfs  computed in  the general
relativity case  crystallize at higher luminosities  (because of their
larger central densities), with the consequent increase in the cooling
times  at those  stages.  In the  1.369$M_{\sun}$ relativistic  sequence,
the  whole
impact of crystallization on the cooling times results smaller, due to
the  fact that  we  neglect  the process  of  phase separation  during
crystallization in that sequence.

We mention  that we neglect the neutrino emission resulting from Urca
process,  which is  relevant  in O/Ne white dwarfs  at  densities in
excess  of $10^{9}$  g cm$^{-3}$ \citep{2021ApJ...916..119S} . In our modeling,  such densities are
attained at  models with  stellar masses $\gtrsim  1.33 M_\sun  $, see
Table  \ref{table1}.  Hence,  the   depicted  cooling  times  for  the
sequences with stellar masses above this value may be overestimated at
high and intermediate  luminosities.  A first attempt  to include Urca
cooling process from $^{23}$Na-$^{23}$Ne urca pair in our stellar code
leads to the  formation of a mixing region below  the Urca shell, as
reported by \cite{2021ApJ...916..119S}. Because of  the
temperature  inversion  caused  by  Urca process, our  most massive
white dwarf models develop off-centered crystallization. 
We  find  numerical
difficulties  to   model  the   interaction  of   crystallization  and
the Urca process-induced mixing  that prevent  us from  a consistent  computation of  white
dwarf   cooling   during   these   stages.  As   recently   shown   by
\cite{2021ApJ...916..119S}, the  cooling of such massive  white dwarfs
is dominated by neutrino cooling via the Urca process during the first
100 Myr after formation.  Our focus in
this work is on the effects of general relativity  on 
ultra-massive  white  dwarfs, so we leave the  problematic  treatment  of
Urca-process  impacts on  the structure of  relativistic white  dwarfs for  an
upcoming work.

%la diferencia es porque el radio es mucho mas pequeño en el caso de relatividad general y como g va
%con 1/R2, eso hace la diferencia. Los efectos de G y V no son importantes pues son proximo a uno

\subsection{Observational constrains on ultra-massive white dwarf models}

\begin{figure*}
  \begin{center}
\subfigure{\includegraphics[width=1.0\columnwidth,clip=true,trim=5 30 55 35]{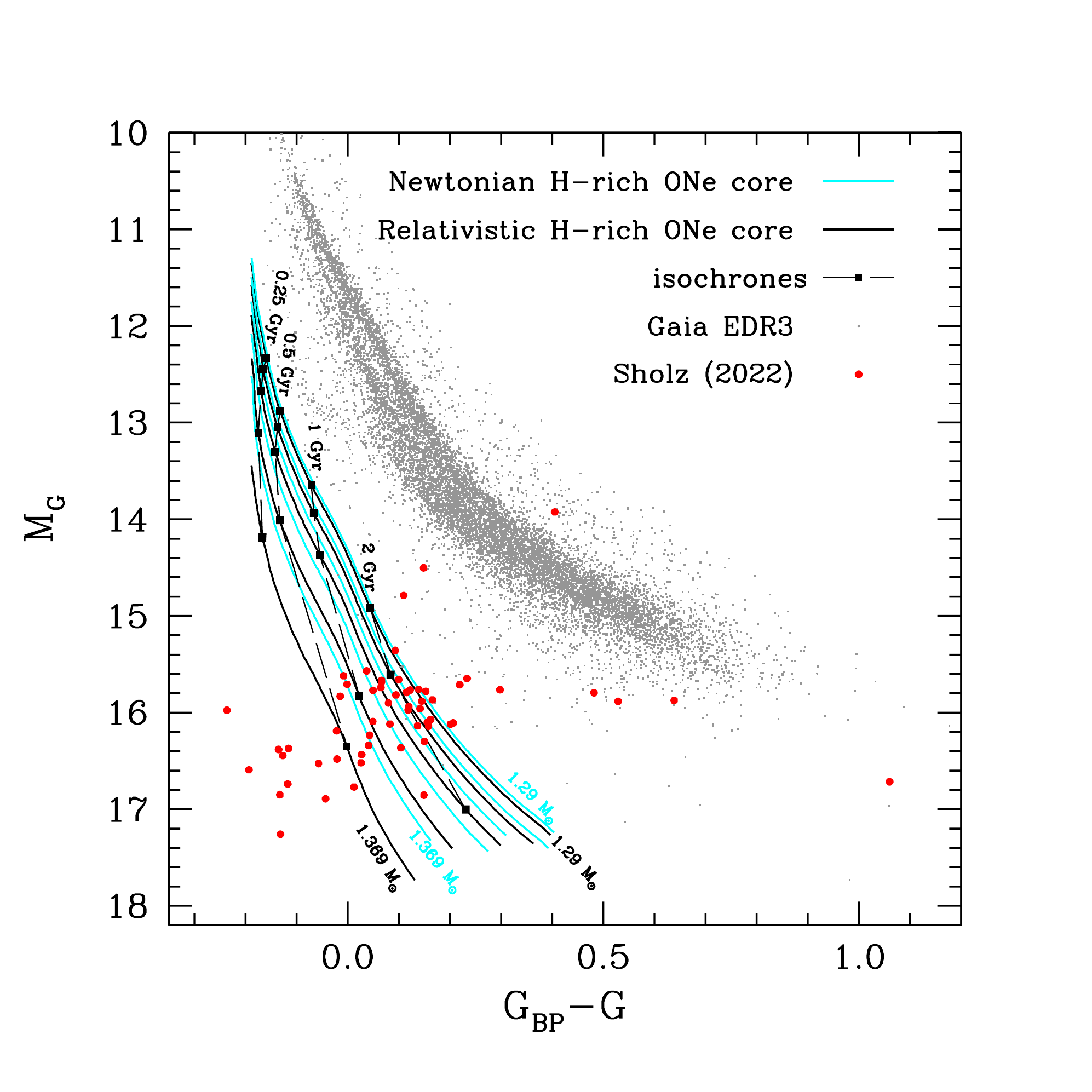}}
\subfigure{\includegraphics[width=1.0\columnwidth,clip=true,trim=5 30 55 35]{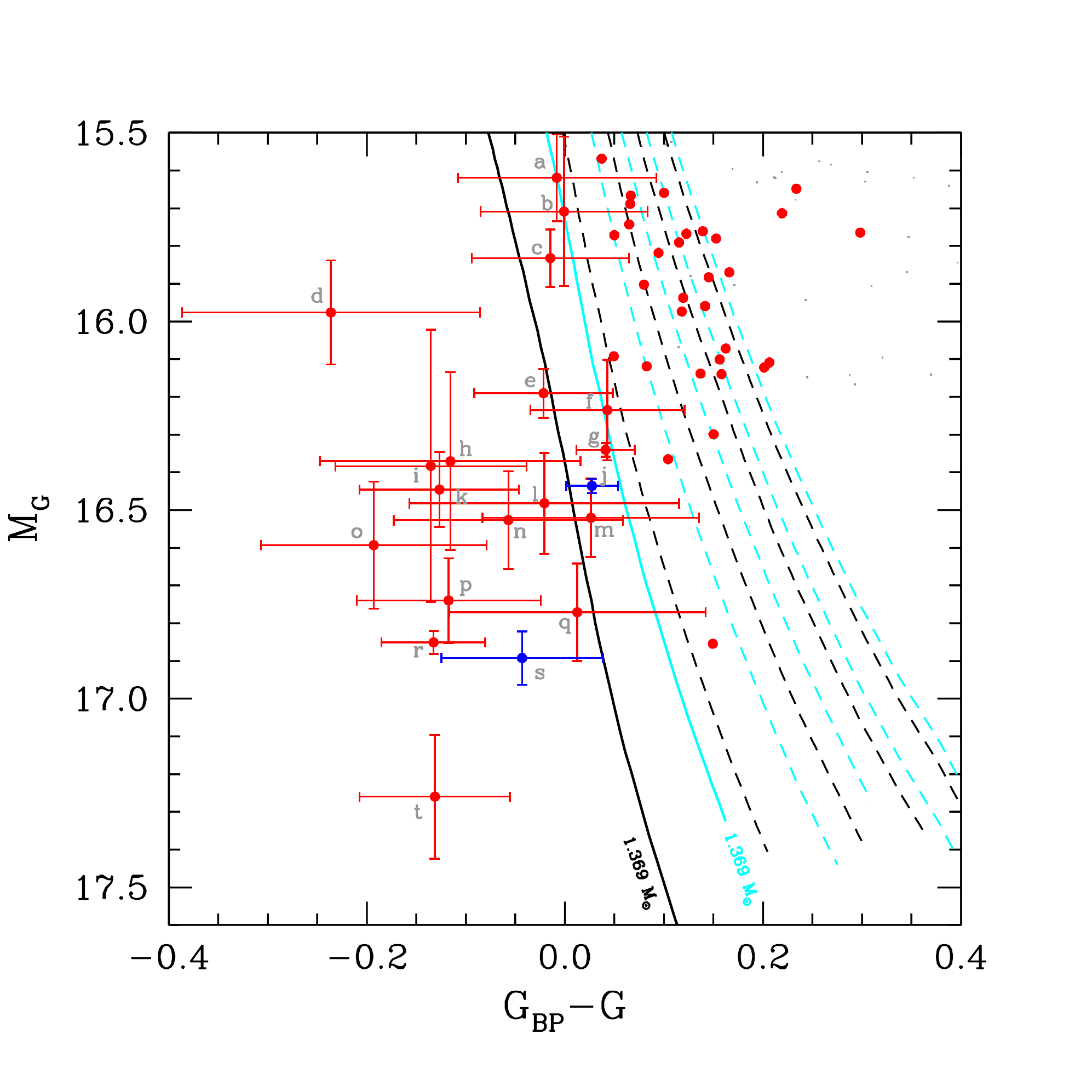}}
  \end{center}
        \caption{{\it Left panel:} {\it Gaia} EDR3 color-magnitude diagram.  Newtonian and general relativistic cooling sequences are displayed using cyan and black lines, respectively. Their rest-masses are, from top to bottom, 1.29, 1.31, 1.33, 1.35, and 1.369$M_{\sun}$. The {\it Gaia} white dwarf population within 100 pc is displayed using gray dots. The faint blue white dwarf branch reported in \cite{2022RNAAS...6...36S} is displayed using large red filled circles. Dashed lines show isochrones of 0.25, 0.5, 1, 2 Gyr.
        {\it Right panel:} Zoomed-in view of the faint blue white dwarf branch. Objects $\{j,s\}$ (marked in blue) are ideal candidates compatible at $1\sigma$ level with the general relativistic model, but marginally at $2\sigma$ level with Newtonian models.} 
        \label{gaia}
\end{figure*}

The ESA  {\it Gaia}  mission has provided  an unprecedented  wealth of
information about stars  \citep[see][and   references  therein]{GaiaEDR32021}.  In
particular, nearly  $\approx$359,000 white dwarf candidates  have been
detected \citep{Fusillo2021},  being estimated  that the sample  up to
100  pc  from  the  Sun  can be  practically  considered  as  complete
\citep{Jimenez2018}.   The  extreme   precision  of   astrometric  and
photometric  measures allow  us to  derive accurate  color-magnitude
diagrams where to test our  models.  Some unexpected peculiar features
have   been  already   observed  in   the  {\it   Gaia}  white   dwarf
color-magnitude diagram \citep{GaiaDR22018}.
%The so called A and B branches, in principle associated to the atmospheric models \citep[e.g.][]{Bergeron2019}, or 
In particular, the Q branch,  due to crystallization and sedimentation
delays,         has          been         extensively         analyzed
\citep{Cheng2019,2019Natur.565..202T,2021A&A...649L...7C}.  However, a
new  branch,   called  faint   blue  branch   has  been   reported  by
\cite{2022RNAAS...6...36S}. This faint blue branch is formed by nearly
$\sim$60  ultracool   and  ultra-massive  objects,  which   have  been
astrometric and photometric verified and cross validated with the {\it
  Gaia} catalogue  of nearby  stars \citep{GaiaNSC2021} and  the white
dwarf  catalogue  of  \cite{Fusillo2021}.   It is  important  also  to
mention that some of these objects  that form this peculiar feature in
the color-magnitude diagram have  already been reported  \cite[][and
  references therein]{Kilic2020}. Most of these white dwarfs exhibit a
near-infrared flux deficit that has  been attributed to the effects of
molecular  collision-induced   absorption  in   mixed  hydrogen-helium
atmospheres, \cite{Bergeron2022}.   Some  issues  still  remain  to  be
clarified   under  this   assumption  and   not  all   the  objects   in
\cite{2022RNAAS...6...36S}   are   present    in   the   analysis   of
\cite{Bergeron2022}.  Consequently, for our purpose here, which is not
in  contradiction with  the analysis  done in  \cite{Bergeron2022}, we
adopted hydrogen-pure atmosphere models for  the analysis of the whole
\cite{2022RNAAS...6...36S}  sample,   where  particular   objects  are
treated individually.
%This faint blue branch is formed by nearly $\sim$60 ultracool and ultra-massive objects which, despite the lack of ultimate spectroscopic confirmation, they have been astrometric and photometric verified and cross validated with the {\it Gaia} catalogue of nearby stars \citep{GaiaNSC2021} and the white dwarf catalogue of \cite{Fusillo2021}.

In the left panel of Fig. \ref{gaia} we show a color-magnitude diagram
for the  100 pc  white dwarf  {\it Gaia}  EDR3 population  (gray dots)
together    with    the    faint     blue    branch    objects    from
\cite{2022RNAAS...6...36S}  (solid red  circles). The  color-magnitude
diagram  selected is  absolute  magnitude $G$  versus $G_{\rm  BP}-$G,
instead  of $G_{\rm  BP}-$G$_{\rm  RP}$, minimizing  in  this way  the
larger errors induced by the $G_{\rm RP}$ filter for faint objects.
%\alb{This explains why some of the white dwarfs appear as outliers in the left panel of Fig. \ref{gaia}}. 
We  also provide  the  magnitudes for  our  relativistic and  Newtonian
models (black and  cyan lines, respectively) in {\it  Gaia} EDR3 passbands (DR2,
Sloan  Digital Sky  Survey, Pan-STARRS  and other  passbands are  also
available under  request) by using the  non-gray model
atmospheres                                                         of
\cite{2010MmSAI..81..921K,2019A&A...628A.102K}.  Isochrones of  0.25,
0.5, 1  and 2, Gyr for  our relativistic model are  also shown (dashed
black  line) in  Fig. \ref{gaia}.  An initial  inspection of  the {\it
  Gaia}  color-magnitude  diagram reveals  that  our  new white  dwarf
sequences are consistent  with most of the  ultra-massive white dwarfs
within 100 pc from the Sun.
%our white dwarf sequences correspond to the most massive white dwarfs within 100 pc from our Sun and, in particular, 
In addition, the  relativistic white dwarf sequences  are fainter than
Newtonian sequences with the  same mass. Therefore, general relativity
effects must be carefully taken into account when determining the mass
and stellar properties  of the most massive white  dwarfs through {\it
  Gaia}  photometry. Not  considering such  effects would  lead to  an
overestimation  of their  mass and  an incorrect  estimation of  their
cooling times. Finally, we check that faint-blue branch objects do not
follow any  particular isochrone,  thus ruling  out a  common temporal
origin of these stars.

A closer look to the faint blue branch is depicted in the right panel of Fig.  \ref{gaia}. The vast majority of faint blue branch white dwarfs appear to have  masses larger than $\sim 1.29\, M_\odot$. Thus, this sample is ideal for testing our models, in particular, those objects which present the largest masses or, equivalently, the smallest radii. Hence, for the analysis presented here and for reasons of completeness we estimated the error bars for those  objects which lie on the left of the Newtonian 1.369 $M_{\sun}$ track. Errors are propagated from the astrometric and photometric errors provided by {\it Gaia} EDR3. Although correlations in {\it Gaia} photometry are very low  we have assumed that some correlation may exist between parameters. This way errors are added linearly and not in quadrature, thus obtaining an upper limit estimate of the error bars.  The parameters corresponding to the 20 selected ultra-massive white dwarf candidates of the faint blue branch are presented in Table \ref{t:FBcandidates}. In the first column we list the {\it Gaia} EDR3 source ID with a label for an easy identification in Fig.  \ref{gaia}. Columns second to fifth present the parallax, apparent and absolute $G$ magnitudes, and color $G_{\rm BP}-G$ with their corresponding error, respectively. Columns sixth and seventh represent the observational distance  within the color-magnitude diagram  measured in $\sigma$ deviations to the limiting 1.369$M_{\sun}$ cooling track when the general relativity model or the Newtonian model, respectively, is used. Finally, the last column is a  5 digits number flag. The first digit indicates if the relative flux error in the G$_\mathrm{BP}$ band is larger or equal to 10\% (1) or smaller (0). %\verb|flux_bp_over_error |$>10$ (0) or $<10$ (1).
The second digit indicates if the relative flux error in the G$_\mathrm{RP}$ band is larger or equal to 10\% (1) or smaller (0). %\verb| flux_rp_over_error |$>10$ (0) or $<10$ (1). 
The third digit indicates if the $\beta$ parameter as defined by \citet{Riello2021} is $\geq 0.1$ (1) or $<0.1$ (0); if 1 then the object is affected by blending. The fourth digit is set to (0) if the renormalized unit vector ruwe \citep{Lindegren2018} is $<1.4$ (indicative that the solution corresponds to a single object) or set to (1) if it is $\geq 1.4$ (bad solution or binary system). The fifth digit indicates if the object passes (1) or not (0) a 5$\sigma$ cut on the corrected G$_\mathrm{BP}$ and G$_\mathrm{RP}$ flux excess ($C^{*}$; \citealt{Riello2021}). %affected by excess factor (1) or not (0) where abs(C*)<=5*sigmac*gmag. 
An ideal case will show a 00000 flag.

The detailed analysis of the color-magnitude distance to the limiting 1.369$M_{\sun}$ relativistic and Newtonian tracks shown in the sixth and seventh columns, respectively, indicates that, on average, the selected faint blue branch objects are more compatible with the general relativistic model than with the Newtonian model. Six of them $\{a,b,c,f,g,m\}$  lie below the limiting 1.369$M_{\sun}$ relativistic track  while they are $1\sigma$ compatible with the Newtonian model. Moreover, up to four objects $\{h,j,n,s\}$ are compatible with the relativistic model at the  $1\sigma$ level, but only marginally at a $2\sigma$ level with the Newtonian model. In particular, objects $\{j,s\}$ are ideal candidates to confirm relativistic models given that they present a 00000 flag, which is indicative of a reliable photometry and astrometry. The rest of objects $\{d,i,k,o,p,r,t\}$ lie at a distance $2\sigma$  or $3\sigma$ (the last two) for the relativistic model, but at larger distances for the Newtonian model (up to $4\sigma$).  According to our study, these objects with such a small radius or larger masses should be unstable against gravitational collapse. However, any conclusion on this should be taken with caution. On one hand,
although some of these objects belong to the sample analyzed by \cite{Bergeron2022} ($d$, J1612$+$5128; $j$, J1251+4403, also named WD1248+443 \citep{Harris2008}; $o$, J1136$-$1057; and $s$, J0416$-$1826) and some near-infrared flux deficit has been reported for them, a more detailed spectroscopic analysis for all of our candidates is deserved for a precise mass and radius estimation.
%these objects present relatively large photometric errors, hence more precise photometric or even spectroscopic confirmation is needed for a precise mass and radius estimation.
%a detailed analysis of the SED of the selected faint blue branch white dwarfs reveals that {\it Gaia} photometry is below PanStarrs photometry, which indicate that the position of most of these objects may move to an upper right location within the color-magnitude diagram, thus implying smaller masses and larger radius. 
On the other hand, the presence of strong internal magnetic fields or a rapid rotation, not considered in this paper, could allow these objects to support the enormous gravity. It has been shown, in the general relativity framework, that including strong magnetic fields and/or a rapid rotation could lead to a smaller radius and/or a larger limiting-mass for the most massive white dwarfs \citep[e.g.][]{2013ApJ...762..117B,2016MNRAS.456.3375B,2015MNRAS.454..752S}. Indeed, the existence of super-Chandrasekhar white dwarfs, with masses $2.1-2.8\,M_\odot$ has been proposed as a possible scenario to explain the  over-luminous Type Ia supernovae SN 2003fg, SN 2006gz, SN 2007if, SN 2009dc (e.g. Howell et al. 2006; Hicken
et al. 2007; Yamanaka et al. 2009; Scalzo et al. 2010; Silverman et al.
2011; Taubenberger et al. 2011). A detailed follow up of these objects is, in any case, deserved and, at the same time, general relativistic models as the ones presented in this work but for white dwarfs with carbon-oxygen cores are expected to play a capital role in the understanding of the true nature of these objects.

\begin{table*}[t]
\centering
\begin{tabular}{cccccccl} 
  \hline
 \hline  \\[-4pt]           
{\it Gaia} EDR3 & $\varpi\pm\sigma_{\varpi}$ & $G\pm\sigma_{\rm G}$ &  $M_{\rm G}\pm\sigma_{M_{\rm G}}$ &  ($G_{\rm BP}-G)\pm\sigma{_{\rm (G_{BP}-G)}}$ & Rel.  &  New. & flags \\
source ID &   (mas) & (mag) & (mag) &  (mag)  &  model   & model &  \\
  \hline \\[-4pt]  
$6565940122868224640^a$	&	$	11.717	\pm	0.592	$	&	$	20.275	\pm	0.005	$	&	$	15.619	\pm	0.115	$	&	$	-0.008	\pm	0.100	$ & $<1$ & 1 & 00100	\\
$1983698716601024512^b$	&	$	10.761	\pm	0.934	$	&	$	20.549	\pm	0.009	$	&	$	15.708	\pm  	0.198	$	&	$	-0.001	\pm	0.084	$ & $<1$ & 1 & 01000		\\
$6211904903507006336^c$	&	$	15.411	\pm	0.501	$	&	$	19.893	\pm	0.006	$	&	$	15.832	\pm	0.076	$	&	$	-0.014	\pm	0.079	$ & $<1$ & 1 & $00000^1$		\\
$1424656526287583744^d$	&	$	11.523	\pm	0.685	$	&	$	20.668	\pm	0.009	$	&	$	15.976	\pm	0.138	$	&	$	-0.236	\pm	0.150	$ & 2 & 2 & $11000^1$		\\
$3585053427252374272^e$	&	$	16.874	\pm	0.464	$	&	$	20.054	\pm	0.005	$	&	$	16.190	\pm	0.065	$	&	$	-0.022	\pm	0.070	$ & 1 & 1 & $01000^1$		\\
$4377579209528621184^f$	&	$	14.828	\pm	0.860	$	&	$	20.379	\pm	0.007	$	&	$	16.235	\pm	0.133	$	&	$	 0.043	\pm	0.078	$ & $<1$ & 1 & 01000		\\
$1505825635741455872^g$	&	$	29.084	\pm	0.190	$	&	$	19.022	\pm	0.004	$	&	$	16.340	\pm	0.018	$	&	$	 0.041	\pm	0.029	$ & $<1$ & 1 & $00100^{1,2}$		\\
$3480787358063803520^h$	&	$	13.189	\pm	1.365	$	&	$	20.769	\pm	0.010	$	&	$	16.370	\pm	0.235	$	&	$	-0.116	\pm	0.132	$ & 1 & 2 & 11000		\\
$4461423190259561728^i$	&	$	12.908	\pm	2.082	$	&	$	20.829	\pm	0.011	$	&	$	16.383	\pm	0.361	$	&	$	-0.135	\pm	0.096	$ & 2 & 2 & 01000		\\

 $ \bf{5064259336725948672^j}$	&	$\bf	30.638	\pm	0.219	$	&	$ \bf	19.005	\pm	0.004	$	&	$ \bf	16.436	\pm	0.019	$	&	$ \bf	0.027	\pm	0.026	$ & \bf{1} & \bf{2} & $\bf{00000}^1$		\\
$534407181320476288^k$	&	$	15.218	\pm	0.640	$	&	$	20.533	\pm	0.008	$	&	$	16.445 	\pm	0.099	$	&	$	-0.127	\pm	0.080	$ & 2 & 3 & $01000$		\\
$5763109404082525696^l$	&	$	16.279	\pm	0.949	$	&	$	20.424	\pm	0.007	$	&	$	16.482	\pm	0.134	$	&	$	-0.021	\pm	0.136	$ & 1 & 1 & $11000^1$		\\
$2858553485723741312^m$	&	$	16.357	\pm	0.715	$	&	$	20.452	\pm	0.009	$	&	$	16.521	\pm	0.104	$	&	$	0.026	\pm	0.109	$ & 1 & 1 & $01000^1$		\\
$6178573689547383168^n$	&	$	17.098	\pm	0.946	$	&	$	20.362	\pm	0.009	$	&	$	16.527	\pm	0.129	$	&	$	-0.057	\pm	0.116	$ & 1 & 2 & $01000^1$		\\
$3586879608689430400^o$	&	$	17.572	\pm	1.299	$	&	$	20.369	\pm	0.007	$	&	$	16.593	\pm	0.168	$	&	$	-0.193	\pm	0.114	$ & 2 & 3 & $01000^1$		\\
$1738863551836243840^p$	&	$	19.444	\pm	0.933	$	&	$	20.296	\pm	0.007	$	&	$	16.740	\pm	0.112	$	&	$	-0.117	\pm	0.093	$ & 2 & 3 & 01000		\\
$6385055135655898496^q$	&	$	16.607	\pm	0.924	$	&	$	20.670	\pm	0.009	$	&	$	16.771	\pm	0.129	$	&	$	0.013	\pm	0.130	$ & 1 & 1 & 11000		\\
$283928743068277376^r$	&	$	27.731	\pm	0.332	$	&	$	19.636	\pm	0.004	$	&	$	16.850	\pm	0.030	$	&	$	-0.133	\pm	0.052	$ & 3 & 4 & 00100		\\
$\bf{1528861748669458432^s}$	&	$	\bf 20.585	\pm	0.614	$	&	$\bf 	20.325	\pm	0.006	$	&	$	\bf 16.892	\pm	0.070	$	&	$	\bf -0.043	\pm	0.082	$	 & \bf 1 & \bf 2 & $\bf{00000}^{1,3}$	\\
$1674805012263764352^t$	&	$	19.661	\pm	1.347	$	&	$	20.792	\pm	0.015	$	&	$	17.260	\pm	0.164	$	&	$	-0.131	\pm	0.076  $    & 3 & 4 & 01000 \\ 
  \hline  
  \end{tabular}                 
  \caption{Ultra-massive white dwarf candidates selected from the sample of faint blue white dwarfs of \cite{2022RNAAS...6...36S}. Sixth and seventh columns indicate the distance within the color-diagram of Fig.  \ref{gaia}  measured in $1\sigma$ deviations form the selected objects to the limiting 1.369 $M_\odot$ cooling tracks for relativistic and Newtonian models, respectively. Objects $j$ and $s$, marked in bold, are ideal candidates with no flags to confirm relativistic models. See text for rest of columns and details. }
    \label{t:FBcandidates}
    \begin{minipage}{\textwidth}
$^{1}$\cite{Bergeron2022}, $^2$\cite{Gates2004} $^3$\cite{Harris2008}
\end{minipage}
\end{table*}

\section{Summary and conclusions}
\label{conclusions}

In this paper, we present the first set of constant rest-mass ultra-massive O/Ne white dwarf cooling tracks with masses  
$M_{\star} > 1.29 M_\sun$, which fully take into account the effects of general relativity on their structural and evolutionary properties. Ultra-massive white dwarfs are relevant in different astrophysical contexts, such as  type Ia supernovae explosions, stellar merger events, and 
the existence of high magnetic field white dwarfs. In addition, they provide insights into the physical processes in the Super Asymptotic Giant Branch phase preceding their formation. In 
the last few years, the existence of such ultra-massive white dwarfs in the solar neighborhood has been reported in several studies, including the recent discover of a branch of faint blue white dwarfs in the color-magnitude diagram \citep{Kilic2020,2022RNAAS...6...36S}. Although some of these objects present an infrared flux deficit, it is also thought to be composed by ultra-massive white dwarfs with masses larger than $1.29\, M_\odot$.
%including the recent discover of a branch of faint blue white dwarfs in the {\it Gaia} color magnitude diagram, which is thought to be composed by ultra-massive white dwarfs with masses larger than $1.29\, M_\odot$ \citep{2022RNAAS...6...36S}. 
It should be noted that shortly, it is very likely that $g$-mode pulsating ultra-massive white dwarfs with masses $M_{\star} \gtrsim 1.29 M_\sun$ will be discovered thanks to space missions such as {\sl TESS} and {\sl Plato} space telescopes, and it will then be possible to study them through asteroseismology.

We have computed the complete evolution of 1.29, 1.31, 1.33, 1.35, and 1.369 $M_{\sun}$ hydrogen-rich white dwarfs  models, assuming an O/Ne composition for
the core. Calculations
have been performed using the La Plata stellar evolution code, {\tt LPCODE}, for which the  
standard equations of stellar structure and evolution have been modified to include the effects of general relativity. To this end, we have  followed the formalism given in \cite{1977ApJ...212..825T}.  Specifically,   the fully general relativistic partial differential equations governing the evolution of a spherically symmetric star are solved in a way they resemble the standard Newtonian equations of stellar structure. For comparison purposes, the same sequences have been computed but for the Newtonian case. Our new white dwarf models include the energy released during the crystallization process, both due to latent heat and the induced chemical redistribution. We provide cooling times and time dependent mass-radius relations for relativistic ultra-massive white dwarfs. We also provide magnitudes in Gaia, Sloan Digital Sky Survey and Pan-STARRS passbands, using the model atmospheres of \cite{2010MmSAI..81..921K,2019A&A...628A.102K}. 
This set of cooling sequences, together with those calculated in \cite{2019A&A...625A..87C} and \cite{2022MNRAS.511.5198C} for lower stellar masses than computed here, provide an appropriate theoretical framework to study the most massive white dwarfs in our Galaxy, superseding all existing calculations of such objects.

As expected, we find that the importance of general relativistic effects increases as the 
stellar mass is increased.  According to our calculations, O/Ne white dwarfs more 
massive than 1.369 $M_{\sun}$ become gravitationally unstable with respect to general relativity effects. When core chemical distribution due to phase separation on crystallization is considered, such instability occurs at somewhat
lower stellar masses, $\gtrsim 1.360  M_\sun $.
For our most massive sequence, the stellar radius becomes 25\% smaller than predicted
by the Newtonian treatment. The evolutionary properties of our ultra-massive white dwarfs are also modified by general relativity effects.  In particular, at advanced stages of evolution, the cooling times for our most massive white dwarf sequence result in about a factor of two shorter than in the Newtonian case. In addition, not considering general relativity  effects when estimating the properties of such objects through photometric and spectroscopic techniques would lead to an overestimation of their mass of 0.015$M_\sun$ near the critical mass.

%\sout{leading to a self-induced termonuclear supernovae.}

We have compared in the color-magnitude diagram our theoretical sequences with the white dwarfs composing the faint blue white dwarf branch \citep{2022RNAAS...6...36S}. We conclude that, regardless the infrared deficit flux that some particular objects may exhibit, several white dwarfs of this branch can present masses larger than $\sim 1.29  M_\sun $ and that it does not coincide with any isochrone nor with any evolutionary track. We found that seven of the white dwarfs in this branch should have a smaller radius than our most massive cooling sequence and should be gravitationally unstable against collapse. However, apart from the need of a more detailed spectroscopic study to accurately characterize the possible effects of the infrared flux deficit in some of these objects, the presence of strong magnetic fields and a rapid rotation, not considered in this study, could favor the stability of such objects, thus supporting the existence of super-Chandrasekhar white dwarfs, that, in the case of CO-core white dwarfs, should likely be the progenitors of the over-luminous Type Ia supernovae SN 2003fg, SN 2006gz, SN 2007if, SN 2009dc. Consequently, a detailed follow-up of these seven objects is required within the framework of the general relativity models exposed here.
\\

As discussed throughout  this work, our new  ultra-massive white dwarf
models for  O/Ne core-chemical  composition constitute  an improvement
over those computed in the  framework of the standard Newtonian theory
of stellar interiors.  Therefore, in support of previous studies, the
effect of general
relativity must be taken into account  to ascertain the true nature of
the  most massive  white  dwarfs, in  particular,  at assessing  their
structural and evolutionary properties.

%
 %We conclude that the effect of general relativity should be
%taken into account at assessing the structural and evolutionary properties of the most massive
%white dwarfs. Our new ultra-massive white dwarf models for 
%O/Ne core-chemical composition provided here constitute an improvement over those computed in the framework of the standard Newtonian theory of stellar interiors. 

%In futures work, we will consider the impact of neutrino emission via Urca process
%and its interaction with both the induced convective core and core crystallization on the structure and evolution of ultra-massive white dwarfs. 
%In addition, we will extend our study to the case of ultra-massive 
%white dwarfs with C/O cores which are relevant for the occurrence of Type Ia supernovae. Finally, we will also explore in future works the pulsational properties of these models to assess the impact of general relativity on $g$-mode pulsation periods and the possible implications for ultra-massive white dwarf asteroseismology in the light of new space observations.

\begin{acknowledgements}

  We thank Detlev Koester for extending his atmosphere models to the high surface gravities that characterize our relativistic ultra-massive white dwarf models.
  We also thank the comments of an anonymous referee that improved the original
  version of this paper.
Part of  this work was  supported by PICT-2017-0884 from ANPCyT, PIP
112-200801-00940 grant from CONICET, grant G149 from University of La Plata, NASA grants 80NSSC17K0008 and 80NSSC20K0193. ST and ARM acknowledge support from MINECO under the PID2020-117252GB-I00 grant. ARM acknowledges support from Grant RYC-2016-20254 funded by MCIN/AEI/10.13039/501100011033 and by ESF Investing in your future. This  research has  made use of  NASA Astrophysics Data System. This work has made use of data from the European Space Agency (ESA) mission {\it Gaia} (\url{https://www.cosmos.esa.int/gaia}), processed by the {\it Gaia} Data Processing and Analysis Consortium (DPAC, \url{https://www.cosmos.esa.int/web/gaia/dpac/consortium}). Funding for the DPAC has been provided by national institutions, in particular the institutions participating in the {\it Gaia} Multilateral Agreement. 
\end{acknowledgements}

\bibliographystyle{aa}
\bibliography{ultramassiveCO}

\begin{thebibliography}{86}
\expandafter\ifx\csname natexlab\endcsname\relax\def\natexlab#1{#1}\fi

\bibitem[{{Althaus} {et~al.}(2015){Althaus}, {Camisassa}, {Miller Bertolami},
  {C{\'o}rsico}, \& {Garc{\'{\i}}a-Berro}}]{2015A&A...576A...9A}
{Althaus}, L.~G., {Camisassa}, M.~E., {Miller Bertolami}, M.~M., {C{\'o}rsico},
  A.~H., \& {Garc{\'{\i}}a-Berro}, E. 2015, \aap, 576, A9

\bibitem[{{Althaus} {et~al.}(2010){Althaus}, {C{\'o}rsico}, {Isern}, \&
  {Garc{\'{\i}}a-Berro}}]{2010A&ARv..18..471A}
{Althaus}, L.~G., {C{\'o}rsico}, A.~H., {Isern}, J., \& {Garc{\'{\i}}a-Berro},
  E. 2010, \aapr, 18, 471

\bibitem[{{Althaus} {et~al.}(2021{\natexlab{a}}){Althaus}, {Gil-Pons},
  {C{\'o}rsico}, {Miller Bertolami}, {De Ger{\'o}nimo}, {Camisassa}, {Torres},
  {Gutierrez}, \& {Rebassa-Mansergas}}]{ALTUMCO2021}
{Althaus}, L.~G., {Gil-Pons}, P., {C{\'o}rsico}, A.~H., {et~al.}
  2021{\natexlab{a}}, \aap, 646, A30

\bibitem[{{Althaus} {et~al.}(2021{\natexlab{b}}){Althaus}, {Gil-Pons},
  {C{\'o}rsico}, {Miller Bertolami}, {De Ger{\'o}nimo}, {Camisassa}, {Torres},
  {Gutierrez}, \& {Rebassa-Mansergas}}]{2021A&A...646A..30A}
---. 2021{\natexlab{b}}, \aap, 646, A30

\bibitem[{{Althaus} {et~al.}(2003){Althaus}, {Serenelli}, {C{\'o}rsico}, \&
  {Montgomery}}]{2003A&A...404..593A}
{Althaus}, L.~G., {Serenelli}, A.~M., {C{\'o}rsico}, A.~H., \& {Montgomery},
  M.~H. 2003, \aap, 404, 593

\bibitem[{{Althaus} {et~al.}(2005){Althaus}, {Serenelli}, {Panei},
  {C{\'o}rsico}, {Garc{\'{\i}}a-Berro}, \&
  {Sc{\'o}ccola}}]{2005A&A...435..631A}
{Althaus}, L.~G., {Serenelli}, A.~M., {Panei}, J.~A., {et~al.} 2005, \aap, 435,
  631

\bibitem[{{Bera} \& {Bhattacharya}(2016)}]{2016MNRAS.456.3375B}
{Bera}, P. \& {Bhattacharya}, D. 2016, \mnras, 456, 3375

\bibitem[{{Bergeron} {et~al.}(2022){Bergeron}, {Kilic}, {Blouin}, {B{\'e}dard},
  {Leggett}, \& {Brown}}]{Bergeron2022}
{Bergeron}, P., {Kilic}, M., {Blouin}, S., {et~al.} 2022, \apj, 934, 36

\bibitem[{{Blouin} \& {Daligault}(2021)}]{2021ApJ...919...87B}
{Blouin}, S. \& {Daligault}, J. 2021, \apj, 919, 87

\bibitem[{{Borucki} {et~al.}(2010){Borucki}, {Koch}, {Basri}, {Batalha},
  {Brown}, {Caldwell}, {Caldwell}, {Christensen-Dalsgaard}, {Cochran},
  {DeVore}, {Dunham}, {Dupree}, {Gautier}, {Geary}, {Gilliland}, {Gould},
  {Howell}, {Jenkins}, {Kondo}, {Latham}, {Marcy}, {Meibom}, {Kjeldsen},
  {Lissauer}, {Monet}, {Morrison}, {Sasselov}, {Tarter}, {Boss}, {Brownlee},
  {Owen}, {Buzasi}, {Charbonneau}, {Doyle}, {Fortney}, {Ford}, {Holman},
  {Seager}, {Steffen}, {Welsh}, {Rowe}, {Anderson}, {Buchhave}, {Ciardi},
  {Walkowicz}, {Sherry}, {Horch}, {Isaacson}, {Everett}, {Fischer}, {Torres},
  {Johnson}, {Endl}, {MacQueen}, {Bryson}, {Dotson}, {Haas}, {Kolodziejczak},
  {Van Cleve}, {Chandrasekaran}, {Twicken}, {Quintana}, {Clarke}, {Allen},
  {Li}, {Wu}, {Tenenbaum}, {Verner}, {Bruhweiler}, {Barnes}, \&
  {Prsa}}]{2010Sci...327..977B}
{Borucki}, W.~J., {Koch}, D., {Basri}, G., {et~al.} 2010, Science, 327, 977

\bibitem[{{Boshkayev} {et~al.}(2013){Boshkayev}, {Rueda}, {Ruffini}, \&
  {Siutsou}}]{2013ApJ...762..117B}
{Boshkayev}, K., {Rueda}, J.~A., {Ruffini}, R., \& {Siutsou}, I. 2013, \apj,
  762, 117

\bibitem[{{Bours} {et~al.}(2015){Bours}, {Marsh}, {G{\"a}nsicke}, {Tauris},
  {Istrate}, {Badenes}, {Dhillon}, {Gal-Yam}, {Hermes}, {Kengkriangkrai},
  {Kilic}, {Koester}, {Mullally}, {Prasert}, {Steeghs}, {Thompson}, \&
  {Thorstensen}}]{2015MNRAS.450.3966B}
{Bours}, M.~C.~P., {Marsh}, T.~R., {G{\"a}nsicke}, B.~T., {et~al.} 2015,
  \mnras, 450, 3966

\bibitem[{{Caiazzo} {et~al.}(2021){Caiazzo}, {Burdge}, {Fuller}, {Heyl},
  {Kulkarni}, {Prince}, {Richer}, {Schwab}, {Andreoni}, {Bellm}, {Drake},
  {Duev}, {Graham}, {Helou}, {Mahabal}, {Masci}, {Smith}, \&
  {Soumagnac}}]{2021Natur.595...39C}
{Caiazzo}, I., {Burdge}, K.~B., {Fuller}, J., {et~al.} 2021, \nat, 595, 39

\bibitem[{{Camisassa} {et~al.}(2019){Camisassa}, {Althaus}, {C{\'o}rsico}, {De
  Ger{\'o}nimo}, {Miller Bertolami}, {Novarino}, {Rohrmann}, {Wachlin}, \&
  {Garc{\'\i}a-Berro}}]{2019A&A...625A..87C}
{Camisassa}, M.~E., {Althaus}, L.~G., {C{\'o}rsico}, A.~H., {et~al.} 2019,
  \aap, 625, A87

\bibitem[{{Camisassa} {et~al.}(2022){Camisassa}, {Althaus}, {Koester},
  {Torres}, {Pons}, \& {C{\'o}rsico}}]{2022MNRAS.511.5198C}
{Camisassa}, M.~E., {Althaus}, L.~G., {Koester}, D., {et~al.} 2022, \mnras,
  511, 5198

\bibitem[{{Camisassa} {et~al.}(2021){Camisassa}, {Althaus}, {Torres},
  {C{\'o}rsico}, {Rebassa-Mansergas}, {Tremblay}, {Cheng}, \&
  {Raddi}}]{2021A&A...649L...7C}
{Camisassa}, M.~E., {Althaus}, L.~G., {Torres}, S., {et~al.} 2021, \aap, 649,
  L7

\bibitem[{{Carvalho} {et~al.}(2018){Carvalho}, {Marinho}, \&
  {Malheiro}}]{2018GReGr..50...38C}
{Carvalho}, G.~A., {Marinho}, R.~M., \& {Malheiro}, M. 2018, General Relativity
  and Gravitation, 50, 38

\bibitem[{{Cheng} {et~al.}(2019){Cheng}, {Cummings}, \&
  {M{\'e}nard}}]{Cheng2019}
{Cheng}, S., {Cummings}, J.~D., \& {M{\'e}nard}, B. 2019, \apj, 886, 100

\bibitem[{{Cheng} {et~al.}(2020){Cheng}, {Cummings}, {M{\'e}nard}, \&
  {Toonen}}]{2020ApJ...891..160C}
{Cheng}, S., {Cummings}, J.~D., {M{\'e}nard}, B., \& {Toonen}, S. 2020, \apj,
  891, 160

\bibitem[{{Christensen-Dalsgaard} {et~al.}(2020){Christensen-Dalsgaard}, {Silva
  Aguirre}, {Cassisi}, {Miller Bertolami}, {Serenelli}, {Stello}, {Weiss},
  {Angelou}, {Jiang}, {Lebreton}, {Spada}, {Bellinger}, {Deheuvels},
  {Ouazzani}, {Pietrinferni}, {Mosumgaard}, {Townsend}, {Battich}, {Bossini},
  {Constantino}, {Eggenberger}, {Hekker}, {Mazumdar}, {Miglio}, {Nielsen}, \&
  {Salaris}}]{2020A&A...635A.165C}
{Christensen-Dalsgaard}, J., {Silva Aguirre}, V., {Cassisi}, S., {et~al.} 2020,
  \aap, 635, A165

\bibitem[{{C{\'o}rsico}(2020)}]{2020FrASS...7...47C}
{C{\'o}rsico}, A.~H. 2020, Frontiers in Astronomy and Space Sciences, 7, 47

\bibitem[{{C{\'o}rsico}(2022)}]{2022arXiv220303769C}
---. 2022, arXiv e-prints, arXiv:2203.03769

\bibitem[{{C{\'o}rsico} {et~al.}(2019{\natexlab{a}}){C{\'o}rsico}, {Althaus},
  {Miller Bertolami}, \& {Kepler}}]{2019A&ARv..27....7C}
{C{\'o}rsico}, A.~H., {Althaus}, L.~G., {Miller Bertolami}, M.~M., \& {Kepler},
  S.~O. 2019{\natexlab{a}}, \aapr, 27, 7

\bibitem[{{C{\'o}rsico} {et~al.}(2019{\natexlab{b}}){C{\'o}rsico}, {De
  Ger{\'o}nimo}, {Camisassa}, \& {Althaus}}]{2019A&A...632A.119C}
{C{\'o}rsico}, A.~H., {De Ger{\'o}nimo}, F.~C., {Camisassa}, M.~E., \&
  {Althaus}, L.~G. 2019{\natexlab{b}}, \aap, 632, A119

\bibitem[{{Curd} {et~al.}(2017){Curd}, {Gianninas}, {Bell}, {Kilic}, {Romero},
  {Allende Prieto}, {Winget}, \& {Winget}}]{2017MNRAS.468..239C}
{Curd}, B., {Gianninas}, A., {Bell}, K.~J., {et~al.} 2017, \mnras, 468, 239

\bibitem[{{de Carvalho} {et~al.}(2014){de Carvalho}, {Rotondo}, {Rueda}, \&
  {Ruffini}}]{2014PhRvC..89a5801D}
{de Carvalho}, S.~M., {Rotondo}, M., {Rueda}, J.~A., \& {Ruffini}, R. 2014,
  \prc, 89, 015801

\bibitem[{{De Ger{\'o}nimo} {et~al.}(2019){De Ger{\'o}nimo}, {C{\'o}rsico},
  {Althaus}, {Wachlin}, \& {Camisassa}}]{2019A&A...621A.100D}
{De Ger{\'o}nimo}, F.~C., {C{\'o}rsico}, A.~H., {Althaus}, L.~G., {Wachlin},
  F.~C., \& {Camisassa}, M.~E. 2019, \aap, 621, A100

\bibitem[{{Doherty} {et~al.}(2010){Doherty}, {Siess}, {Lattanzio}, \&
  {Gil-Pons}}]{2010MNRAS.401.1453D}
{Doherty}, C.~L., {Siess}, L., {Lattanzio}, J.~C., \& {Gil-Pons}, P. 2010,
  \mnras, 401, 1453

\bibitem[{{Gagn{\'e}} {et~al.}(2018){Gagn{\'e}}, {Fontaine}, {Simon}, \&
  {Faherty}}]{2018ApJ...861L..13G}
{Gagn{\'e}}, J., {Fontaine}, G., {Simon}, A., \& {Faherty}, J.~K. 2018, \apjl,
  861, L13

\bibitem[{{Gaia Collaboration} {et~al.}(2018){Gaia Collaboration}, {Babusiaux},
  {van Leeuwen}, {Barstow}, {Jordi}, {Vallenari}, {Bossini}, {Bressan},
  {Cantat-Gaudin}, {van Leeuwen}, {Brown}, {Prusti}, {de Bruijne},
  {Bailer-Jones}, {Biermann}, {Evans}, {Eyer}, {Jansen}, {Klioner}, {Lammers},
  {Lindegren}, {Luri}, {Mignard}, {Panem}, {Pourbaix}, {Randich}, {Sartoretti},
  {Siddiqui}, {Soubiran}, {Walton}, {Arenou}, {Bastian}, {Cropper}, {Drimmel},
  {Katz}, {Lattanzi}, {Bakker}, {Cacciari}, {Casta{\~n}eda}, {Chaoul}, {Cheek},
  {De Angeli}, {Fabricius}, {Guerra}, {Holl}, {Masana}, {Messineo}, {Mowlavi},
  {Nienartowicz}, {Panuzzo}, {Portell}, {Riello}, {Seabroke}, {Tanga},
  {Th{\'e}venin}, {Gracia-Abril}, {Comoretto}, {Garcia-Reinaldos}, {Teyssier},
  {Altmann}, {Andrae}, {Audard}, {Bellas-Velidis}, {Benson}, {Berthier},
  {Blomme}, {Burgess}, {Busso}, {Carry}, {Cellino}, {Clementini}, {Clotet},
  {Creevey}, {Davidson}, {De Ridder}, {Delchambre}, {Dell'Oro}, {Ducourant},
  {Fern{\'a}ndez-Hern{\'a}ndez}, {Fouesneau}, {Fr{\'e}mat}, {Galluccio},
  {Garc{\'\i}a-Torres}, {Gonz{\'a}lez-N{\'u}{\~n}ez}, {Gonz{\'a}lez-Vidal},
  {Gosset}, {Guy}, {Halbwachs}, {Hambly}, {Harrison}, {Hern{\'a}ndez},
  {Hestroffer}, {Hodgkin}, {Hutton}, {Jasniewicz}, {Jean-Antoine-Piccolo},
  {Jordan}, {Korn}, {Krone-Martins}, {Lanzafame}, {Lebzelter}, {L{\"o}ffler},
  {Manteiga}, {Marrese}, {Mart{\'\i}n-Fleitas}, {Moitinho}, {Mora}, {Muinonen},
  {Osinde}, {Pancino}, {Pauwels}, {Petit}, {Recio-Blanco}, {Richards},
  {Rimoldini}, {Robin}, {Sarro}, {Siopis}, {Smith}, {Sozzetti}, {S{\"u}veges},
  {Torra}, {van Reeven}, {Abbas}, {Abreu Aramburu}, {Accart}, {Aerts},
  {Altavilla}, {{\'A}lvarez}, {Alvarez}, {Alves}, {Anderson}, {Andrei},
  {Anglada Varela}, {Antiche}, {Antoja}, {Arcay}, {Astraatmadja}, {Bach},
  {Baker}, {Balaguer-N{\'u}{\~n}ez}, {Balm}, {Barache}, {Barata}, {Barbato},
  {Barblan}, {Barklem}, {Barrado}, {Barros}, {Bartholom{\'e} Mu{\~n}oz},
  {Bassilana}, {Becciani}, {Bellazzini}, {Berihuete}, {Bertone}, {Bianchi},
  {Bienaym{\'e}}, {Blanco-Cuaresma}, {Boch}, {Boeche}, {Bombrun}, {Borrachero},
  {Bouquillon}, {Bourda}, {Bragaglia}, {Bramante}, {Breddels}, {Brouillet},
  {Br{\"u}semeister}, {Brugaletta}, {Bucciarelli}, {Burlacu}, {Busonero},
  {Butkevich}, {Buzzi}, {Caffau}, {Cancelliere}, {Cannizzaro}, {Carballo},
  {Carlucci}, {Carrasco}, {Casamiquela}, {Castellani}, {Castro-Ginard},
  {Charlot}, {Chemin}, {Chiavassa}, {Cocozza}, {Costigan}, {Cowell}, {Crifo},
  {Crosta}, {Crowley}, {Cuypers}, {Dafonte}, {Damerdji}, {Dapergolas}, {David},
  {David}, {de Laverny}, {De Luise}, {De March}, {de Martino}, {de Souza}, {de
  Torres}, {Debosscher}, {del Pozo}, {Delbo}, {Delgado}, {Delgado}, {Diakite},
  {Diener}, {Distefano}, {Dolding}, {Drazinos}, {Dur{\'a}n}, {Edvardsson},
  {Enke}, {Eriksson}, {Esquej}, {Eynard Bontemps}, {Fabre}, {Fabrizio},
  {Faigler}, {Falc{\~a}o}, {Farr{\`a}s Casas}, {Federici}, {Fedorets},
  {Fernique}, {Figueras}, {Filippi}, {Findeisen}, {Fonti}, {Fraile}, {Fraser},
  {Fr{\'e}zouls}, {Gai}, {Galleti}, {Garabato}, {Garc{\'\i}a-Sedano},
  {Garofalo}, {Garralda}, {Gavel}, {Gavras}, {Gerssen}, {Geyer}, {Giacobbe},
  {Gilmore}, {Girona}, {Giuffrida}, {Glass}, {Gomes}, {Granvik}, {Gueguen},
  {Guerrier}, {Guiraud}, {Guti{\'e}}, {Haigron}, {Hatzidimitriou}, {Hauser},
  {Haywood}, {Heiter}, {Helmi}, {Heu}, {Hilger}, {Hobbs}, {Hofmann}, {Holland},
  {Huckle}, {Hypki}, {Icardi}, {Jan{\ss}en}, {Jevardat de Fombelle}, {Jonker},
  {Juh{\'a}sz}, {Julbe}, {Karampelas}, {Kewley}, {Klar}, {Kochoska}, {Kohley},
  {Kolenberg}, {Kontizas}, {Kontizas}, {Koposov}, {Kordopatis},
  {Kostrzewa-Rutkowska}, {Koubsky}, {Lambert}, {Lanza}, {Lasne}, {Lavigne}, {Le
  Fustec}, {Le Poncin-Lafitte}, {Lebreton}, {Leccia}, {Leclerc},
  {Lecoeur-Taibi}, {Lenhardt}, {Leroux}, {Liao}, {Licata}, {Lindstr{\o}m},
  {Lister}, {Livanou}, {Lobel}, {L{\'o}pez}, {Managau}, {Mann}, {Mantelet},
  {Marchal}, {Marchant}, {Marconi}, {Marinoni}, {Marschalk{\'o}}, {Marshall},
  {Martino}, {Marton}, {Mary}, {Massari}, {Matijevi{\v{c}}}, {Mazeh},
  {McMillan}, {Messina}, {Michalik}, {Millar}, {Molina}, {Molinaro},
  {Moln{\'a}r}, {Montegriffo}, {Mor}, {Morbidelli}, {Morel}, {Morris},
  {Mulone}, {Muraveva}, {Musella}, {Nelemans}, {Nicastro}, {Noval},
  {O'Mullane}, {Ord{\'e}novic}, {Ord{\'o}{\~n}ez-Blanco}, {Osborne}, {Pagani},
  {Pagano}, {Pailler}, {Palacin}, {Palaversa}, {Panahi}, {Pawlak},
  {Piersimoni}, {Pineau}, {Plachy}, {Plum}, {Poggio}, {Poujoulet},
  {Pr{\v{s}}a}, {Pulone}, {Racero}, {Ragaini}, {Rambaux}, {Ramos-Lerate},
  {Regibo}, {Reyl{\'e}}, {Riclet}, {Ripepi}, {Riva}, {Rivard}, {Rixon},
  {Roegiers}, {Roelens}, {Romero-G{\'o}mez}, {Rowell}, {Royer}, {Ruiz-Dern},
  {Sadowski}, {Sagrist{\`a} Sell{\'e}s}, {Sahlmann}, {Salgado}, {Salguero},
  {Sanna}, {Santana-Ros}, {Sarasso}, {Savietto}, {Schultheis}, {Sciacca},
  {Segol}, {Segovia}, {S{\'e}gransan}, {Shih}, {Siltala}, {Silva}, {Smart},
  {Smith}, {Solano}, {Solitro}, {Sordo}, {Soria Nieto}, {Souchay}, {Spagna},
  {Spoto}, {Stampa}, {Steele}, {Steidelm{\"u}ller}, {Stephenson}, {Stoev},
  {Suess}, {Surdej}, {Szabados}, {Szegedi-Elek}, {Tapiador}, {Taris}, {Tauran},
  {Taylor}, {Teixeira}, {Terrett}, {Teyssandier}, {Thuillot}, {Titarenko},
  {Torra Clotet}, {Turon}, {Ulla}, {Utrilla}, {Uzzi}, {Vaillant}, {Valentini},
  {Valette}, {van Elteren}, {Van Hemelryck}, {Vaschetto}, {Vecchiato},
  {Veljanoski}, {Viala}, {Vicente}, {Vogt}, {von Essen}, {Voss}, {Votruba},
  {Voutsinas}, {Walmsley}, {Weiler}, {Wertz}, {Wevers}, {Wyrzykowski},
  {Yoldas}, {{\v{Z}}erjal}, {Ziaeepour}, {Zorec}, {Zschocke}, {Zucker},
  {Zurbach}, \& {Zwitter}}]{GaiaDR22018}
{Gaia Collaboration}, {Babusiaux}, C., {van Leeuwen}, F., {et~al.} 2018, \aap,
  616, A10

\bibitem[{{Gaia Collaboration} {et~al.}(2021{\natexlab{a}}){Gaia
  Collaboration}, {Brown}, {Vallenari}, {Prusti}, {de Bruijne}, {Babusiaux},
  {Biermann}, {Creevey}, {Evans}, {Eyer}, {Hutton}, {Jansen}, {Jordi},
  {Klioner}, {Lammers}, {Lindegren}, {Luri}, {Mignard}, {Panem}, {Pourbaix},
  {Randich}, {Sartoretti}, {Soubiran}, {Walton}, {Arenou}, {Bailer-Jones},
  {Bastian}, {Cropper}, {Drimmel}, {Katz}, {Lattanzi}, {van Leeuwen}, {Bakker},
  {Cacciari}, {Casta{\~n}eda}, {De Angeli}, {Ducourant}, {Fabricius},
  {Fouesneau}, {Fr{\'e}mat}, {Guerra}, {Guerrier}, {Guiraud}, {Jean-Antoine
  Piccolo}, {Masana}, {Messineo}, {Mowlavi}, {Nicolas}, {Nienartowicz},
  {Pailler}, {Panuzzo}, {Riclet}, {Roux}, {Seabroke}, {Sordo}, {Tanga},
  {Th{\'e}venin}, {Gracia-Abril}, {Portell}, {Teyssier}, {Altmann}, {Andrae},
  {Bellas-Velidis}, {Benson}, {Berthier}, {Blomme}, {Brugaletta}, {Burgess},
  {Busso}, {Carry}, {Cellino}, {Cheek}, {Clementini}, {Damerdji}, {Davidson},
  {Delchambre}, {Dell'Oro}, {Fern{\'a}ndez-Hern{\'a}ndez}, {Galluccio},
  {Garc{\'\i}a-Lario}, {Garcia-Reinaldos}, {Gonz{\'a}lez-N{\'u}{\~n}ez},
  {Gosset}, {Haigron}, {Halbwachs}, {Hambly}, {Harrison}, {Hatzidimitriou},
  {Heiter}, {Hern{\'a}ndez}, {Hestroffer}, {Hodgkin}, {Holl}, {Jan{\ss}en},
  {Jevardat de Fombelle}, {Jordan}, {Krone-Martins}, {Lanzafame},
  {L{\"o}ffler}, {Lorca}, {Manteiga}, {Marchal}, {Marrese}, {Moitinho}, {Mora},
  {Muinonen}, {Osborne}, {Pancino}, {Pauwels}, {Petit}, {Recio-Blanco},
  {Richards}, {Riello}, {Rimoldini}, {Robin}, {Roegiers}, {Rybizki}, {Sarro},
  {Siopis}, {Smith}, {Sozzetti}, {Ulla}, {Utrilla}, {van Leeuwen}, {van
  Reeven}, {Abbas}, {Abreu Aramburu}, {Accart}, {Aerts}, {Aguado}, {Ajaj},
  {Altavilla}, {{\'A}lvarez}, {{\'A}lvarez Cid-Fuentes}, {Alves}, {Anderson},
  {Anglada Varela}, {Antoja}, {Audard}, {Baines}, {Baker},
  {Balaguer-N{\'u}{\~n}ez}, {Balbinot}, {Balog}, {Barache}, {Barbato},
  {Barros}, {Barstow}, {Bartolom{\'e}}, {Bassilana}, {Bauchet},
  {Baudesson-Stella}, {Becciani}, {Bellazzini}, {Bernet}, {Bertone}, {Bianchi},
  {Blanco-Cuaresma}, {Boch}, {Bombrun}, {Bossini}, {Bouquillon}, {Bragaglia},
  {Bramante}, {Breedt}, {Bressan}, {Brouillet}, {Bucciarelli}, {Burlacu},
  {Busonero}, {Butkevich}, {Buzzi}, {Caffau}, {Cancelliere}, {C{\'a}novas},
  {Cantat-Gaudin}, {Carballo}, {Carlucci}, {Carnerero}, {Carrasco},
  {Casamiquela}, {Castellani}, {Castro-Ginard}, {Castro Sampol}, {Chaoul},
  {Charlot}, {Chemin}, {Chiavassa}, {Cioni}, {Comoretto}, {Cooper}, {Cornez},
  {Cowell}, {Crifo}, {Crosta}, {Crowley}, {Dafonte}, {Dapergolas}, {David},
  {David}, {de Laverny}, {De Luise}, {De March}, {De Ridder}, {de Souza}, {de
  Teodoro}, {de Torres}, {del Peloso}, {del Pozo}, {Delbo}, {Delgado},
  {Delgado}, {Delisle}, {Di Matteo}, {Diakite}, {Diener}, {Distefano},
  {Dolding}, {Eappachen}, {Edvardsson}, {Enke}, {Esquej}, {Fabre}, {Fabrizio},
  {Faigler}, {Fedorets}, {Fernique}, {Fienga}, {Figueras}, {Fouron},
  {Fragkoudi}, {Fraile}, {Franke}, {Gai}, {Garabato}, {Garcia-Gutierrez},
  {Garc{\'\i}a-Torres}, {Garofalo}, {Gavras}, {Gerlach}, {Geyer}, {Giacobbe},
  {Gilmore}, {Girona}, {Giuffrida}, {Gomel}, {Gomez}, {Gonzalez-Santamaria},
  {Gonz{\'a}lez-Vidal}, {Granvik}, {Guti{\'e}rrez-S{\'a}nchez}, {Guy},
  {Hauser}, {Haywood}, {Helmi}, {Hidalgo}, {Hilger}, {H{\l}adczuk}, {Hobbs},
  {Holland}, {Huckle}, {Jasniewicz}, {Jonker}, {Juaristi Campillo}, {Julbe},
  {Karbevska}, {Kervella}, {Khanna}, {Kochoska}, {Kontizas}, {Kordopatis},
  {Korn}, {Kostrzewa-Rutkowska}, {Kruszy{\'n}ska}, {Lambert}, {Lanza}, {Lasne},
  {Le Campion}, {Le Fustec}, {Lebreton}, {Lebzelter}, {Leccia}, {Leclerc},
  {Lecoeur-Taibi}, {Liao}, {Licata}, {Lindstr{\o}m}, {Lister}, {Livanou},
  {Lobel}, {Madrero Pardo}, {Managau}, {Mann}, {Marchant}, {Marconi}, {Marcos
  Santos}, {Marinoni}, {Marocco}, {Marshall}, {Martin Polo},
  {Mart{\'\i}n-Fleitas}, {Masip}, {Massari}, {Mastrobuono-Battisti}, {Mazeh},
  {McMillan}, {Messina}, {Michalik}, {Millar}, {Mints}, {Molina}, {Molinaro},
  {Moln{\'a}r}, {Montegriffo}, {Mor}, {Morbidelli}, {Morel}, {Morris},
  {Mulone}, {Munoz}, {Muraveva}, {Murphy}, {Musella}, {Noval}, {Ord{\'e}novic},
  {Orr{\`u}}, {Osinde}, {Pagani}, {Pagano}, {Palaversa}, {Palicio}, {Panahi},
  {Pawlak}, {Pe{\~n}alosa Esteller}, {Penttil{\"a}}, {Piersimoni}, {Pineau},
  {Plachy}, {Plum}, {Poggio}, {Poretti}, {Poujoulet}, {Pr{\v{s}}a}, {Pulone},
  {Racero}, {Ragaini}, {Rainer}, {Raiteri}, {Rambaux}, {Ramos}, {Ramos-Lerate},
  {Re Fiorentin}, {Regibo}, {Reyl{\'e}}, {Ripepi}, {Riva}, {Rixon}, {Robichon},
  {Robin}, {Roelens}, {Rohrbasser}, {Romero-G{\'o}mez}, {Rowell}, {Royer},
  {Rybicki}, {Sadowski}, {Sagrist{\`a} Sell{\'e}s}, {Sahlmann}, {Salgado},
  {Salguero}, {Samaras}, {Sanchez Gimenez}, {Sanna}, {Santove{\~n}a},
  {Sarasso}, {Schultheis}, {Sciacca}, {Segol}, {Segovia}, {S{\'e}gransan},
  {Semeux}, {Shahaf}, {Siddiqui}, {Siebert}, {Siltala}, {Slezak}, {Smart},
  {Solano}, {Solitro}, {Souami}, {Souchay}, {Spagna}, {Spoto}, {Steele},
  {Steidelm{\"u}ller}, {Stephenson}, {S{\"u}veges}, {Szabados}, {Szegedi-Elek},
  {Taris}, {Tauran}, {Taylor}, {Teixeira}, {Thuillot}, {Tonello}, {Torra},
  {Torra}, {Turon}, {Unger}, {Vaillant}, {van Dillen}, {Vanel}, {Vecchiato},
  {Viala}, {Vicente}, {Voutsinas}, {Weiler}, {Wevers}, {Wyrzykowski}, {Yoldas},
  {Yvard}, {Zhao}, {Zorec}, {Zucker}, {Zurbach}, \& {Zwitter}}]{GaiaEDR32021}
{Gaia Collaboration}, {Brown}, A.~G.~A., {Vallenari}, A., {et~al.}
  2021{\natexlab{a}}, \aap, 649, A1

\bibitem[{{Gaia Collaboration} {et~al.}(2021{\natexlab{b}}){Gaia
  Collaboration}, {Smart}, {Sarro}, {Rybizki}, {Reyl{\'e}}, {Robin}, {Hambly},
  {Abbas}, {Barstow}, {de Bruijne}, {Bucciarelli}, {Carrasco}, {Cooper},
  {Hodgkin}, {Masana}, {Michalik}, {Sahlmann}, {Sozzetti}, {Brown},
  {Vallenari}, {Prusti}, {Babusiaux}, {Biermann}, {Creevey}, {Evans}, {Eyer},
  {Hutton}, {Jansen}, {Jordi}, {Klioner}, {Lammers}, {Lindegren}, {Luri},
  {Mignard}, {Panem}, {Pourbaix}, {Randich}, {Sartoretti}, {Soubiran},
  {Walton}, {Arenou}, {Bailer-Jones}, {Bastian}, {Cropper}, {Drimmel}, {Katz},
  {Lattanzi}, {van Leeuwen}, {Bakker}, {Casta{\~n}eda}, {De Angeli},
  {Ducourant}, {Fabricius}, {Fouesneau}, {Fr{\'e}mat}, {Guerra}, {Guerrier},
  {Guiraud}, {Jean-Antoine Piccolo}, {Messineo}, {Mowlavi}, {Nicolas},
  {Nienartowicz}, {Pailler}, {Panuzzo}, {Riclet}, {Roux}, {Seabroke}, {Sordo},
  {Tanga}, {Th{\'e}venin}, {Gracia-Abril}, {Portell}, {Teyssier}, {Altmann},
  {Andrae}, {Bellas-Velidis}, {Benson}, {Berthier}, {Blomme}, {Brugaletta},
  {Burgess}, {Busso}, {Carry}, {Cellino}, {Cheek}, {Clementini}, {Damerdji},
  {Davidson}, {Delchambre}, {Dell'Oro}, {Fern{\'a}ndez-Hern{\'a}ndez},
  {Galluccio}, {Garc{\'\i}a-Lario}, {Garcia-Reinaldos},
  {Gonz{\'a}lez-N{\'u}{\~n}ez}, {Gosset}, {Haigron}, {Halbwachs}, {Harrison},
  {Hatzidimitriou}, {Heiter}, {Hern{\'a}ndez}, {Hestroffer}, {Holl},
  {Jan{\ss}en}, {Jevardat de Fombelle}, {Jordan}, {Krone-Martins}, {Lanzafame},
  {L{\"o}ffler}, {Lorca}, {Manteiga}, {Marchal}, {Marrese}, {Moitinho}, {Mora},
  {Muinonen}, {Osborne}, {Pancino}, {Pauwels}, {Recio-Blanco}, {Richards},
  {Riello}, {Rimoldini}, {Roegiers}, {Siopis}, {Smith}, {Ulla}, {Utrilla}, {van
  Leeuwen}, {van Reeven}, {Abreu Aramburu}, {Accart}, {Aerts}, {Aguado},
  {Ajaj}, {Altavilla}, {{\'A}lvarez}, {{\'A}lvarez Cid-Fuentes}, {Alves},
  {Anderson}, {Anglada Varela}, {Antoja}, {Audard}, {Baines}, {Baker},
  {Balaguer-N{\'u}{\~n}ez}, {Balbinot}, {Balog}, {Barache}, {Barbato},
  {Barros}, {Bartolom{\'e}}, {Bassilana}, {Bauchet}, {Baudesson-Stella},
  {Becciani}, {Bellazzini}, {Bernet}, {Bertone}, {Bianchi}, {Blanco-Cuaresma},
  {Boch}, {Bombrun}, {Bossini}, {Bouquillon}, {Bragaglia}, {Bramante},
  {Breedt}, {Bressan}, {Brouillet}, {Burlacu}, {Busonero}, {Butkevich},
  {Buzzi}, {Caffau}, {Cancelliere}, {C{\'a}novas}, {Cantat-Gaudin}, {Carballo},
  {Carlucci}, {Carnerero}, {Casamiquela}, {Castellani}, {Castro-Ginard},
  {Castro Sampol}, {Chaoul}, {Charlot}, {Chemin}, {Chiavassa}, {Cioni},
  {Comoretto}, {Cornez}, {Cowell}, {Crifo}, {Crosta}, {Crowley}, {Dafonte},
  {Dapergolas}, {David}, {David}, {de Laverny}, {De Luise}, {De March}, {De
  Ridder}, {de Souza}, {de Teodoro}, {de Torres}, {del Peloso}, {del Pozo},
  {Delgado}, {Delgado}, {Delisle}, {Di Matteo}, {Diakite}, {Diener},
  {Distefano}, {Dolding}, {Eappachen}, {Edvardsson}, {Enke}, {Esquej}, {Fabre},
  {Fabrizio}, {Faigler}, {Fedorets}, {Fernique}, {Fienga}, {Figueras},
  {Fouron}, {Fragkoudi}, {Fraile}, {Franke}, {Gai}, {Garabato},
  {Garcia-Gutierrez}, {Garc{\'\i}a-Torres}, {Garofalo}, {Gavras}, {Gerlach},
  {Geyer}, {Giacobbe}, {Gilmore}, {Girona}, {Giuffrida}, {Gomel}, {Gomez},
  {Gonzalez-Santamaria}, {Gonz{\'a}lez-Vidal}, {Granvik},
  {Guti{\'e}rrez-S{\'a}nchez}, {Guy}, {Hauser}, {Haywood}, {Helmi}, {Hidalgo},
  {Hilger}, {H{\l}adczuk}, {Hobbs}, {Holland}, {Huckle}, {Jasniewicz},
  {Jonker}, {Juaristi Campillo}, {Julbe}, {Karbevska}, {Kervella}, {Khanna},
  {Kochoska}, {Kontizas}, {Kordopatis}, {Korn}, {Kostrzewa-Rutkowska},
  {Kruszy{\'n}ska}, {Lambert}, {Lanza}, {Lasne}, {Le Campion}, {Le Fustec},
  {Lebreton}, {Lebzelter}, {Leccia}, {Leclerc}, {Lecoeur-Taibi}, {Liao},
  {Licata}, {Lindstr{\o}m}, {Lister}, {Livanou}, {Lobel}, {Madrero Pardo},
  {Managau}, {Mann}, {Marchant}, {Marconi}, {Marcos Santos}, {Marinoni},
  {Marocco}, {Marshall}, {Martin Polo}, {Mart{\'\i}n-Fleitas}, {Masip},
  {Massari}, {Mastrobuono-Battisti}, {Mazeh}, {McMillan}, {Messina}, {Millar},
  {Mints}, {Molina}, {Molinaro}, {Moln{\'a}r}, {Montegriffo}, {Mor},
  {Morbidelli}, {Morel}, {Morris}, {Mulone}, {Munoz}, {Muraveva}, {Murphy},
  {Musella}, {Noval}, {Ord{\'e}novic}, {Orr{\`u}}, {Osinde}, {Pagani},
  {Pagano}, {Palaversa}, {Palicio}, {Panahi}, {Pawlak}, {Pe{\~n}alosa
  Esteller}, {Penttil{\"a}}, {Piersimoni}, {Pineau}, {Plachy}, {Plum},
  {Poggio}, {Poretti}, {Poujoulet}, {Pr{\v{s}}a}, {Pulone}, {Racero},
  {Ragaini}, {Rainer}, {Raiteri}, {Rambaux}, {Ramos}, {Ramos-Lerate}, {Re
  Fiorentin}, {Regibo}, {Ripepi}, {Riva}, {Rixon}, {Robichon}, {Robin},
  {Roelens}, {Rohrbasser}, {Romero-G{\'o}mez}, {Rowell}, {Royer}, {Rybicki},
  {Sadowski}, {Sagrist{\`a} Sell{\'e}s}, {Salgado}, {Salguero}, {Samaras},
  {Sanchez Gimenez}, {Sanna}, {Santove{\~n}a}, {Sarasso}, {Schultheis},
  {Sciacca}, {Segol}, {Segovia}, {S{\'e}gransan}, {Semeux}, {Shahaf},
  {Siddiqui}, {Siebert}, {Siltala}, {Slezak}, {Solano}, {Solitro}, {Souami},
  {Souchay}, {Spagna}, {Spoto}, {Steele}, {Steidelm{\"u}ller}, {Stephenson},
  {S{\"u}veges}, {Szabados}, {Szegedi-Elek}, {Taris}, {Tauran}, {Taylor},
  {Teixeira}, {Thuillot}, {Tonello}, {Torra}, {Torra}, {Turon}, {Unger},
  {Vaillant}, {van Dillen}, {Vanel}, {Vecchiato}, {Viala}, {Vicente},
  {Voutsinas}, {Weiler}, {Wevers}, {Wyrzykowski}, {Yoldas}, {Yvard}, {Zhao},
  {Zorec}, {Zucker}, {Zurbach}, \& {Zwitter}}]{GaiaNSC2021}
{Gaia Collaboration}, {Smart}, R.~L., {Sarro}, L.~M., {et~al.}
  2021{\natexlab{b}}, \aap, 649, A6

\bibitem[{{Garc{\'{\i}}a-Berro} \& {Oswalt}(2016)}]{2016NewAR..72....1G}
{Garc{\'{\i}}a-Berro}, E. \& {Oswalt}, T.~D. 2016, New Astronomy Reviews, 72, 1

\bibitem[{{Garc{\'\i}a-Berro} {et~al.}(1997){Garc{\'\i}a-Berro}, {Ritossa}, \&
  {Iben}}]{1997ApJ...485..765G}
{Garc{\'\i}a-Berro}, E., {Ritossa}, C., \& {Iben}, Icko, J. 1997, \apj, 485,
  765

\bibitem[{{Gates} {et~al.}(2004){Gates}, {Gyuk}, {Harris}, {Subbarao},
  {Anderson}, {Kleinman}, {Liebert}, {Brewington}, {Brinkmann}, {Harvanek},
  {Krzesinski}, {Lamb}, {Long}, {Neilsen}, {Newman}, {Nitta}, \&
  {Snedden}}]{Gates2004}
{Gates}, E., {Gyuk}, G., {Harris}, H.~C., {et~al.} 2004, \apjl, 612, L129

\bibitem[{{Gentile Fusillo} {et~al.}(2021){Gentile Fusillo}, {Tremblay},
  {Cukanovaite}, {Vorontseva}, {Lallement}, {Hollands}, {G{\"a}nsicke},
  {Burdge}, {McCleery}, \& {Jordan}}]{Fusillo2021}
{Gentile Fusillo}, N.~P., {Tremblay}, P.~E., {Cukanovaite}, E., {et~al.} 2021,
  \mnras, 508, 3877

\bibitem[{{Gianninas} {et~al.}(2011){Gianninas}, {Bergeron}, \&
  {Ruiz}}]{2011ApJ...743..138G}
{Gianninas}, A., {Bergeron}, P., \& {Ruiz}, M.~T. 2011, \apj, 743, 138

\bibitem[{{Gil-Pons} {et~al.}(2005){Gil-Pons}, {Suda}, {Fujimoto}, \&
  {Garc{\'\i}a-Berro}}]{2005A&A...433.1037G}
{Gil-Pons}, P., {Suda}, T., {Fujimoto}, M.~Y., \& {Garc{\'\i}a-Berro}, E. 2005,
  \aap, 433, 1037

\bibitem[{{Haft} {et~al.}(1994){Haft}, {Raffelt}, \&
  {Weiss}}]{1994ApJ...425..222H}
{Haft}, M., {Raffelt}, G., \& {Weiss}, A. 1994, \apj, 425, 222

\bibitem[{{Harris} {et~al.}(2008){Harris}, {Gates}, {Gyuk}, {Subbarao},
  {Anderson}, {Hall}, {Munn}, {Liebert}, {Knapp}, {Bizyaev}, {Malanushenko},
  {Malanushenko}, {Pan}, {Schneider}, \& {Allyn Smith}}]{Harris2008}
{Harris}, H.~C., {Gates}, E., {Gyuk}, G., {et~al.} 2008, \apj, 679, 697

\bibitem[{{Hermes} {et~al.}(2013){Hermes}, {Kepler}, {Castanheira},
  {Gianninas}, {Winget}, {Montgomery}, {Brown}, \&
  {Harrold}}]{2013ApJ...771L...2H}
{Hermes}, J.~J., {Kepler}, S.~O., {Castanheira}, B.~G., {et~al.} 2013, \apjl,
  771, L2

\bibitem[{{Hollands} {et~al.}(2020){Hollands}, {Tremblay}, {G{\"a}nsicke},
  {Camisassa}, {Koester}, {Aungwerojwit}, {Chote}, {C{\'o}rsico}, {Dhillon},
  {Gentile-Fusillo}, {Hoskin}, {Izquierdo}, {Marsh}, \&
  {Steeghs}}]{Hollands2020}
{Hollands}, M.~A., {Tremblay}, P.~E., {G{\"a}nsicke}, B.~T., {et~al.} 2020,
  Nature Astronomy

\bibitem[{{Howell} {et~al.}(2014){Howell}, {Sobeck}, {Haas}, {Still},
  {Barclay}, {Mullally}, {Troeltzsch}, {Aigrain}, {Bryson}, {Caldwell},
  {Chaplin}, {Cochran}, {Huber}, {Marcy}, {Miglio}, {Najita}, {Smith},
  {Twicken}, \& {Fortney}}]{2014PASP..126..398H}
{Howell}, S.~B., {Sobeck}, C., {Haas}, M., {et~al.} 2014, \pasp, 126, 398

\bibitem[{{Isern} {et~al.}(2022){Isern}, {Torres}, \&
  {Rebassa-Mansergas}}]{2022FrASS...9....6I}
{Isern}, J., {Torres}, S., \& {Rebassa-Mansergas}, A. 2022, Frontiers in
  Astronomy and Space Sciences, 9, 6

\bibitem[{{Itoh} {et~al.}(1996){Itoh}, {Hayashi}, {Nishikawa}, \&
  {Kohyama}}]{1996ApJS..102..411I}
{Itoh}, N., {Hayashi}, H., {Nishikawa}, A., \& {Kohyama}, Y. 1996, ApJs, 102,
  411

\bibitem[{{Jim{\'e}nez-Esteban} {et~al.}(2018){Jim{\'e}nez-Esteban}, {Torres},
  {Rebassa-Mansergas}, {Skorobogatov}, {Solano}, {Cantero}, \&
  {Rodrigo}}]{Jimenez2018}
{Jim{\'e}nez-Esteban}, F.~M., {Torres}, S., {Rebassa-Mansergas}, A., {et~al.}
  2018, \mnras, 480, 4505

\bibitem[{{Kanaan} {et~al.}(1992){Kanaan}, {Kepler}, {Giovannini}, \&
  {Diaz}}]{1992ApJ...390L..89K}
{Kanaan}, A., {Kepler}, S.~O., {Giovannini}, O., \& {Diaz}, M. 1992, \apjl,
  390, L89

\bibitem[{{Kepler} {et~al.}(2016){Kepler}, {Pelisoli}, {Koester}, {Ourique},
  {Romero}, {Reindl}, {Kleinman}, {Eisenstein}, {Valois}, \&
  {Amaral}}]{2016MNRAS.455.3413K}
{Kepler}, S.~O., {Pelisoli}, I., {Koester}, D., {et~al.} 2016, \mnras, 455,
  3413

\bibitem[{{Kilic} {et~al.}(2021){Kilic}, {Bergeron}, {Blouin}, \&
  {B{\'e}dard}}]{2021MNRAS.503.5397K}
{Kilic}, M., {Bergeron}, P., {Blouin}, S., \& {B{\'e}dard}, A. 2021, \mnras,
  503, 5397

\bibitem[{{Kilic} {et~al.}(2020){Kilic}, {Bergeron}, {Kosakowski}, {Brown},
  {Ag{\"u}eros}, \& {Blouin}}]{Kilic2020}
{Kilic}, M., {Bergeron}, P., {Kosakowski}, A., {et~al.} 2020, \apj, 898, 84

\bibitem[{{Kippenhahn} {et~al.}(2012){Kippenhahn}, {Weigert}, \&
  {Weiss}}]{2012sse..book.....K}
{Kippenhahn}, R., {Weigert}, A., \& {Weiss}, A. 2012, {Stellar Structure and
  Evolution}

\bibitem[{{Kleinman} {et~al.}(2013){Kleinman}, {Kepler}, {Koester}, {Pelisoli},
  {Pe{\c c}anha}, {Nitta}, {Costa}, {Krzesinski}, {Dufour}, {Lachapelle},
  {Bergeron}, {Yip}, {Harris}, {Eisenstein}, {Althaus}, \&
  {C{\'o}rsico}}]{2013ApJS..204....5K}
{Kleinman}, S.~J., {Kepler}, S.~O., {Koester}, D., {et~al.} 2013, ApJs, 204, 5

\bibitem[{{Koester}(2010)}]{2010MmSAI..81..921K}
{Koester}, D. 2010, \memsai, 81, 921

\bibitem[{{Koester} \& {Kepler}(2019)}]{2019A&A...628A.102K}
{Koester}, D. \& {Kepler}, S.~O. 2019, \aap, 628, A102

\bibitem[{{Lindegren} {et~al.}(2018){Lindegren}, {Hern{\'a}ndez}, {Bombrun},
  {Klioner}, {Bastian}, {Ramos-Lerate}, {de Torres}, {Steidelm{\"u}ller},
  {Stephenson}, {Hobbs}, {Lammers}, {Biermann}, {Geyer}, {Hilger}, {Michalik},
  {Stampa}, {McMillan}, {Casta{\~n}eda}, {Clotet}, {Comoretto}, {Davidson},
  {Fabricius}, {Gracia}, {Hambly}, {Hutton}, {Mora}, {Portell}, {van Leeuwen},
  {Abbas}, {Abreu}, {Altmann}, {Andrei}, {Anglada}, {Balaguer-N{\'u}{\~n}ez},
  {Barache}, {Becciani}, {Bertone}, {Bianchi}, {Bouquillon}, {Bourda},
  {Br{\"u}semeister}, {Bucciarelli}, {Busonero}, {Buzzi}, {Cancelliere},
  {Carlucci}, {Charlot}, {Cheek}, {Crosta}, {Crowley}, {de Bruijne}, {de
  Felice}, {Drimmel}, {Esquej}, {Fienga}, {Fraile}, {Gai}, {Garralda},
  {Gonz{\'a}lez-Vidal}, {Guerra}, {Hauser}, {Hofmann}, {Holl}, {Jordan},
  {Lattanzi}, {Lenhardt}, {Liao}, {Licata}, {Lister}, {L{\"o}ffler},
  {Marchant}, {Martin-Fleitas}, {Messineo}, {Mignard}, {Morbidelli}, {Poggio},
  {Riva}, {Rowell}, {Salguero}, {Sarasso}, {Sciacca}, {Siddiqui}, {Smart},
  {Spagna}, {Steele}, {Taris}, {Torra}, {van Elteren}, {van Reeven}, \&
  {Vecchiato}}]{Lindegren2018}
{Lindegren}, L., {Hern{\'a}ndez}, J., {Bombrun}, A., {et~al.} 2018, \aap, 616,
  A2

\bibitem[{{Magni} \& {Mazzitelli}(1979)}]{1979A&A....72..134M}
{Magni}, G. \& {Mazzitelli}, I. 1979, \aap, 72, 134

\bibitem[{{Mathew} \& {Nandy}(2017)}]{2017RAA....17...61M}
{Mathew}, A. \& {Nandy}, M.~K. 2017, Research in Astronomy and Astrophysics,
  17, 061

\bibitem[{{Medin} \& {Cumming}(2010)}]{2010PhRvE..81c6107M}
{Medin}, Z. \& {Cumming}, A. 2010, \pre, 81, 036107

\bibitem[{{Miller Bertolami}(2016)}]{2016A&A...588A..25M}
{Miller Bertolami}, M.~M. 2016, \aap, 588, A25

\bibitem[{{Moriya}(2019)}]{2019MNRAS.490.1166M}
{Moriya}, T.~J. 2019, \mnras, 490, 1166

\bibitem[{{Moya} {et~al.}(2018){Moya}, {Barcel{\'o} Forteza}, {Bonfanti},
  {Salmon}, {Van Grootel}, \& {Barrado}}]{2018A&A...620A.203M}
{Moya}, A., {Barcel{\'o} Forteza}, S., {Bonfanti}, A., {et~al.} 2018, \aap,
  620, A203

\bibitem[{{Mukadam} {et~al.}(2004){Mukadam}, {Mullally}, {Nather}, {Winget},
  {von Hippel}, {Kleinman}, {Nitta}, {Krzesi{\'n}ski}, {Kepler}, {Kanaan},
  {Koester}, {Sullivan}, {Homeier}, {Thompson}, {Reaves}, {Cotter},
  {Slaughter}, \& {Brinkmann}}]{2004ApJ...607..982M}
{Mukadam}, A.~S., {Mullally}, F., {Nather}, R.~E., {et~al.} 2004, \apj, 607,
  982

\bibitem[{{Nitta} {et~al.}(2016){Nitta}, {Kepler}, {Chen{\'e}}, {Koester},
  {Provencal}, {Kleinmani}, {Sullivan}, {Chote}, {Sefako}, {Kanaan}, {Romero},
  {Corti}, {Kilic}, {Montgomery}, \& {Winget}}]{2016IAUFM..29B.493N}
{Nitta}, A., {Kepler}, S.~O., {Chen{\'e}}, A.-N., {et~al.} 2016, IAU Focus
  Meeting, 29, 493

\bibitem[{{Nunes} {et~al.}(2021){Nunes}, {Arba{\~n}il}, \&
  {Malheiro}}]{2021ApJ...921..138N}
{Nunes}, S.~P., {Arba{\~n}il}, J. D.~V., \& {Malheiro}, M. 2021, \apj, 921, 138

\bibitem[{{Piotto}(2018)}]{2018EPSC...12..969P}
{Piotto}, G. 2018, in European Planetary Science Congress, EPSC2018--969

\bibitem[{{Pshirkov} {et~al.}(2020){Pshirkov}, {Dodin}, {Belinski},
  {Zheltoukhov}, {Fedoteva}, {Voziakova}, {Potanin}, {Blinnikov}, \&
  {Postnov}}]{2020MNRAS.499L..21P}
{Pshirkov}, M.~S., {Dodin}, A.~V., {Belinski}, A.~A., {et~al.} 2020, \mnras,
  499, L21

\bibitem[{{Ricker} {et~al.}(2015){Ricker}, {Winn}, {Vanderspek}, {Latham},
  {Bakos}, {Bean}, {Berta-Thompson}, {Brown}, {Buchhave}, {Butler}, {Butler},
  {Chaplin}, {Charbonneau}, {Christensen-Dalsgaard}, {Clampin}, {Deming},
  {Doty}, {De Lee}, {Dressing}, {Dunham}, {Endl}, {Fressin}, {Ge}, {Henning},
  {Holman}, {Howard}, {Ida}, {Jenkins}, {Jernigan}, {Johnson}, {Kaltenegger},
  {Kawai}, {Kjeldsen}, {Laughlin}, {Levine}, {Lin}, {Lissauer}, {MacQueen},
  {Marcy}, {McCullough}, {Morton}, {Narita}, {Paegert}, {Palle}, {Pepe},
  {Pepper}, {Quirrenbach}, {Rinehart}, {Sasselov}, {Sato}, {Seager},
  {Sozzetti}, {Stassun}, {Sullivan}, {Szentgyorgyi}, {Torres}, {Udry}, \&
  {Villasenor}}]{2015JATIS...1a4003R}
{Ricker}, G.~R., {Winn}, J.~N., {Vanderspek}, R., {et~al.} 2015, Journal of
  Astronomical Telescopes, Instruments, and Systems, 1, 014003

\bibitem[{{Riello} {et~al.}(2021){Riello}, {De Angeli}, {Evans}, {Montegriffo},
  {Carrasco}, {Busso}, {Palaversa}, {Burgess}, {Diener}, {Davidson}, {Rowell},
  {Fabricius}, {Jordi}, {Bellazzini}, {Pancino}, {Harrison}, {Cacciari}, {van
  Leeuwen}, {Hambly}, {Hodgkin}, {Osborne}, {Altavilla}, {Barstow}, {Brown},
  {Castellani}, {Cowell}, {De Luise}, {Gilmore}, {Giuffrida}, {Hidalgo},
  {Holland}, {Marinoni}, {Pagani}, {Piersimoni}, {Pulone}, {Ragaini}, {Rainer},
  {Richards}, {Sanna}, {Walton}, {Weiler}, \& {Yoldas}}]{Riello2021}
{Riello}, M., {De Angeli}, F., {Evans}, D.~W., {et~al.} 2021, \aap, 649, A3

\bibitem[{{Rotondo} {et~al.}(2011){Rotondo}, {Rueda}, {Ruffini}, \&
  {Xue}}]{2011PhRvD..84h4007R}
{Rotondo}, M., {Rueda}, J.~A., {Ruffini}, R., \& {Xue}, S.-S. 2011, \prd, 84,
  084007

\bibitem[{{Rowan} {et~al.}(2019){Rowan}, {Tucker}, {Shappee}, \&
  {Hermes}}]{2019MNRAS.486.4574R}
{Rowan}, D.~M., {Tucker}, M.~A., {Shappee}, B.~J., \& {Hermes}, J.~J. 2019,
  \mnras, 486, 4574

\bibitem[{{Salaris} {et~al.}(2013){Salaris}, {Althaus}, \&
  {Garc{\'\i}a-Berro}}]{2013A&A...555A..96S}
{Salaris}, M., {Althaus}, L.~G., \& {Garc{\'\i}a-Berro}, E. 2013, \aap, 555,
  A96

\bibitem[{{Scholz}(2022)}]{2022RNAAS...6...36S}
{Scholz}, R.-D. 2022, Research Notes of the American Astronomical Society, 6,
  36

\bibitem[{{Schwab}(2021{\natexlab{a}})}]{2021ApJ...916..119S}
{Schwab}, J. 2021{\natexlab{a}}, \apj, 916, 119

\bibitem[{{Schwab}(2021{\natexlab{b}})}]{2021ApJ...906...53S}
---. 2021{\natexlab{b}}, \apj, 906, 53

\bibitem[{{Segretain} {et~al.}(1994){Segretain}, {Chabrier}, {Hernanz},
  {Garcia-Berro}, {Isern}, \& {Mochkovitch}}]{1994ApJ...434..641S}
{Segretain}, L., {Chabrier}, G., {Hernanz}, M., {et~al.} 1994, \apj, 434, 641

\bibitem[{{Siess}(2006)}]{2006A&A...448..717S}
{Siess}, L. 2006, \aap, 448, 717

\bibitem[{{Siess}(2010)}]{2010A&A...512A..10S}
---. 2010, \aap, 512, A10

\bibitem[{{Silva Aguirre} {et~al.}(2020){Silva Aguirre},
  {Christensen-Dalsgaard}, {Cassisi}, {Miller Bertolami}, {Serenelli},
  {Stello}, {Weiss}, {Angelou}, {Jiang}, {Lebreton}, {Spada}, {Bellinger},
  {Deheuvels}, {Ouazzani}, {Pietrinferni}, {Mosumgaard}, {Townsend}, {Battich},
  {Bossini}, {Constantino}, {Eggenberger}, {Hekker}, {Mazumdar}, {Miglio},
  {Nielsen}, \& {Salaris}}]{2020A&A...635A.164S}
{Silva Aguirre}, V., {Christensen-Dalsgaard}, J., {Cassisi}, S., {et~al.} 2020,
  \aap, 635, A164

\bibitem[{{Subramanian} \& {Mukhopadhyay}(2015)}]{2015MNRAS.454..752S}
{Subramanian}, S. \& {Mukhopadhyay}, B. 2015, \mnras, 454, 752

\bibitem[{{Temmink} {et~al.}(2020){Temmink}, {Toonen}, {Zapartas}, {Justham},
  \& {G{\"a}nsicke}}]{2020A&A...636A..31T}
{Temmink}, K.~D., {Toonen}, S., {Zapartas}, E., {Justham}, S., \&
  {G{\"a}nsicke}, B.~T. 2020, \aap, 636, A31

\bibitem[{{Thorne}(1977)}]{1977ApJ...212..825T}
{Thorne}, K.~S. 1977, \apj, 212, 825

\bibitem[{{Torres} {et~al.}(2022){Torres}, {Canals}, {Jim{\'e}nez-Esteban},
  {Rebassa-Mansergas}, \& {Solano}}]{2022MNRAS.511.5462T}
{Torres}, S., {Canals}, P., {Jim{\'e}nez-Esteban}, F.~M., {Rebassa-Mansergas},
  A., \& {Solano}, E. 2022, \mnras, 511, 5462

\bibitem[{{Tremblay} {et~al.}(2019){Tremblay}, {Fontaine}, {Fusillo}, {Dunlap},
  {G{\"a}nsicke}, {Hollands}, {Hermes}, {Marsh}, {Cukanovaite}, \&
  {Cunningham}}]{2019Natur.565..202T}
{Tremblay}, P.-E., {Fontaine}, G., {Fusillo}, N.~P.~G., {et~al.} 2019, \nat,
  565, 202

\bibitem[{{Ventura} \& {D'Antona}(2011)}]{2011MNRAS.410.2760V}
{Ventura}, P. \& {D'Antona}, F. 2011, \mnras, 410, 2760

\bibitem[{{Winget} \& {Kepler}(2008)}]{2008ARA&A..46..157W}
{Winget}, D.~E. \& {Kepler}, S.~O. 2008, \araa, 46, 157

\bibitem[{{Wu} {et~al.}(2022){Wu}, {Xiong}, \& {Wang}}]{2022arXiv220202040W}
{Wu}, C., {Xiong}, H., \& {Wang}, X. 2022, \mnras, 512, 2972

\end{thebibliography}

%\newpage

%\begin{appendix}

\end{document}